\documentclass[english]{article}
\usepackage[T1]{fontenc}
\usepackage[latin9]{inputenc}
\usepackage{booktabs}
\usepackage{units}
\usepackage{amsmath}
\usepackage{amsthm}
\usepackage{amssymb}
\usepackage{amsmath}
\usepackage{float} 
\usepackage{placeins}
\usepackage{color}
\usepackage{array}
\usepackage{float}
\usepackage{booktabs}
\usepackage{multirow}
\usepackage{amsmath}
\usepackage{hyperref}
\usepackage{amsmath}
\usepackage{algorithm,algorithmic}

\DeclareMathOperator*{\argmin}{arg\,min}
\makeatletter

\providecommand{\tabularnewline}{\\}

\numberwithin{equation}{section}
\numberwithin{figure}{section}
\theoremstyle{plain}
\newtheorem{thm}{\protect\theoremname}
\theoremstyle{remark}
\newtheorem{rem}[thm]{\protect\remarkname}

\usepackage{babel}
\usepackage{fullpage}
\usepackage{authblk}
\usepackage{hyperref}
\usepackage{placeins}
\usepackage{hyphenat} 
\usepackage{graphicx}
\usepackage{amsthm}
\usepackage{fancyvrb}
\usepackage[hypcap]{caption}
\usepackage[title,titletoc]{appendix}
\addto\captionsenglish{}

\title{Backward Hedging for American Options with Transaction Costs}
\author{ \textsc{Ludovic Goudenège}\thanks{F\'ederation de Math\'ematiques de CentraleSupelec - CNRS FR3487, France -\texttt{ ludovic.goudenege@math.cnrs.fr}} 
	\and \textsc{Andrea Molent}   \thanks{Dipartimento di Scienze Economiche e Statistiche, Universit\`a degli Studi di Udine, Italy - \texttt{andrea.molent@uniud.it}} 
	\and \textsc{Antonino Zanette}\thanks{Dipartimento di Scienze Economiche e Statistiche, Universit\`a degli Studi di Udine, Italy - \texttt{antonino.zanette@uniud.it}}}
\date{}

\definecolor{azure}{rgb}{0.122, 0.435, 0.698}
\definecolor{arancio}{RGB}{244,70,20}
\definecolor{verde}{RGB}{0,204,20}
\definecolor{blu}{RGB}{20,70,240}

\numberwithin{equation}{section}
\numberwithin{figure}{section}

\makeatother

\usepackage{babel}
\providecommand{\remarkname}{Remark}
\providecommand{\theoremname}{Theorem}

\begin{document}
	\maketitle
	
	\begin{flushleft}
		\rule{1\columnwidth}{1pt}
		\par\end{flushleft}
	
	\begin{flushleft}
		\textbf{\large{}Abstract}{\large\par}
		\par\end{flushleft}
	
	\noindent In this article, we introduce an algorithm called Backward
	Hedging, designed for hedging European and American options while
	considering transaction costs. The optimal strategy is determined
	by minimizing an appropriate loss function, which is based on either
	a risk measure or the mean squared error of the hedging strategy at
	maturity. {The proposed algorithm moves backward in time, determining, for each time-step and different market states, the optimal hedging strategy that minimizes the loss function at the time the option is exercised, by assuming that the strategy used in the future for hedging the liability is the one determined at the previous steps of the algorithm.} The approach
	avoids machine learning and instead relies on {classic} optimization
	techniques, Monte Carlo simulations, and interpolations on a grid.
	Comparisons with the Deep Hedging algorithm in various numerical experiments
	showcase the efficiency and accuracy of the proposed method. 
	
	\vspace{2mm}
	
	\noindent \emph{\large{}Keywords}: Hedging, Greeks, Transaction costs,
	American options, Heston model 
	
	\noindent \rule{1\columnwidth}{1pt}
	
	\newpage
	
	\section{Introduction}
	The groundbreaking work {of} Black and Scholes \cite{black1973pricing}, followed by Merton \cite{merton1973theory}, enabled the objective determination of {the price of} a derivative product as the monetary amount required {by an agent} for the exact replication of its payoff. 
	This definition assumes that exact replication is achievable, aiming to eliminate any potential risk from holding a short position on the derivative. To achieve this, the authors make several assumptions, the most apparent being the possibility of continuously rebalancing the hedging portfolio. This assumption, however, is not feasible in practice due to physical constraints, making it necessary to rebalance the portfolio at { a set of } discrete times. 
	These considerations also apply to advanced models, which incorporate additional risk factors, such as Heston's stochastic volatility model \cite{heston1993closed}. The infeasibility of continuous-time hedging causes these models to transform into approximate versions of themselves, which are associated with incomplete markets: exact replication of derivative products is generally not possible. 
	
	In real-world applications, practitioners typically use continuous-time models to calculate the hedging strategy, despite the impossibility of continuous-time hedging. This is done to exploit the computational power of such models and, specifically, to derive hedging strategies. Furthermore, they include a suitable margin in the price of the derivative to account for the replication error due to discrete-time hedging based on a continuous-time model.
	
	Another limitation to the use of standard theoretical models in a practical context is the presence of frictions, primarily the transaction costs that arise during the construction and rebalancing of the {hedging} portfolio. Naturally, these costs, while making the replication procedure more expensive on one hand, also discourage the rebalancing {prescribed} by the theoretical model on the other. Consequently, it is essential to strike a balance in order to minimize the overall cost of the strategy without significantly compromising its effectiveness.
	
	The first studies related to a pricing model in a imperfect market were developed by Duffie and Richardson \cite{duffie1991mean}. They introduce the mean-variance hedging problem in continuous time, when the hedger can only trade futures contracts on an asset which is positively correlated with the underlying and {so they} consider the problem of  determining the optimal hedge. Schweizer \cite{schweizer1992mean, schweizer1995variance} extends the results of Duffie and Richardson and obtains the first ground-breaking result by solving the hedging problem for a general contingent claim when a mean-squared error loss function is considered. Subsequently, Follmer  \cite{follmer1999quantile, follmer2000efficient} studies theoretical properties of hedging strategies when the loss to minimize is given by the VaR or the CVaR at maturity, computed from the perspective of the hedger. Other authors focus on the development of numerical algorithms for computing optimal hedging strategies. For example, Potter et al. \cite{potters2001hedged} develop an algorithm based on Monte Carlo simulations to determine a hedging strategy that minimizes the quadratic error of hedging between two rebalancing steps. Pochart and Bouchaud \cite{pochart2004option} continue this analysis, by considering minimization of a local risk function such as Var and CVaR, also including transaction costs through a first order approximation.
	The use of convex risk measures for option pricing has been studied also by other authors. For example, Xu \cite{xu2006risk} investigates from a theoretical perspective the evaluation of financial option by risk measure pricing rules. Also Kl{\"o}ppel and Schweizer \cite{kloppel2007dynamic} develop a theoretical investigation and find interesting results on time-consistency about families of conditional convex risk measures.
	To the same scope, some authors study asymptotic behaviour of prices and strategies. With this regard, Whalley and Wilmott \cite{whalley1997asymptotic} develop a model for pricing European options in the presence of rehedging transaction costs by means of asymptotic analysis. 
	Kallsen and Muhle-Karbe  \cite{kallsen2015option} obtain asymptotic formulas for utility indifference prices and hedging strategies in the presence of small transaction costs. Similarly, Bouchard et al. \cite{bouchard2016hedging} obtain explicit formulae for option hedging in a market with proportional transaction costs.

	Recently, some authors have proposed to use modern reinforcement
	machine learning methods  to compute an optimal hedging strategy through neural networks. One of the main results in this sense is due to Buehler et al. \cite{buehler2019deep}. They address the problem of hedging under conditions of incomplete markets with transaction costs by proposing an algorithm   called Deep Hedging: the optimal strategy is determined by training a complex neural network so that it approximates an optimal hedging strategy in terms of an appropriate risk measure. Their approach is greek--free in that the hedging strategy is not determined from the greeks, but rather as the minimizer of a loss function, that is as a solution to a variational problem. The resulting algorithm  for finding the optimal hedging strategy has many merits: it is model free, it can manage transaction costs, it can also include, in the feature set, additional information such as trading signals, news analytics or past hedging decisions. 
	{This new algorithm has enjoyed considerable success in the financial landscape and beyond. For example,  Horvat et al \cite{horvath2021deep} investigate the hedging of short positions on options under rough volatility models again by using the Deep Hedging algorithm. Furthermore, the Deep Hedging algorithm has  been recently employed in the insurance field by Carbonneau \cite{carbonneau2021deep}, with the scope of hedging long term insurance products such as variable annuities.
	}
	
	Despite its merits, there are also some elements that limit the use of the Deep Hedging algorithm: in order to obtain reasonable results, the calculation time required is generally high as the proposed neural network optimises the whole hedging strategy over the entire historical series passed as input. Training such a complex neural network requires considerable numerical effort, and sometimes many iterations are required to converge to the optimal strategy. 	Moreover, neural networks behave as a black box, which may meet with resistance from regulators {(see e.g. Lin \cite{lin2019artificial})}. The Deep Hedging algorithm, as it stands, is specifically designed for European options and {it is not suitable for handling} American options. \\ In the context of American options, we would like to mention the work by Becker et al. \cite{becker2020pricing}, which employs neural networks for pricing and hedging these options.

	In this paper, {motivated by} the {algorithm} introduced by Buehler et al. \cite{buehler2019deep}, we propose an alternative approach that avoids the use of machine learning and relies on traditional optimization and interpolation techniques. We construct an appropriate discretization of the possible market states by using a grid of values, and our algorithm moves backward in time to compute the optimal strategy from each individual state. Consequently, we term this the \textit{Backward Hedging} (BH) algorithm. The optimal strategy is determined by calculating the optimal quantities of hedging instruments to include in the portfolio, minimizing the desired loss function when applied to forward Monte Carlo simulations generated from the state under consideration. Our proposed method proves to be efficient and reliable, with easily interpretable results. {We stress out that, because of lack of time-consistency for most of risk measures (see e.g. Boda and Filar \cite{boda2006time}, Cheridito and Stadje \cite{cheridito2009time}, Cui et al. \cite{cui2019time}), the problem solved by the Backward Hedging algorithm is not equivalent  to the global one, which is solved by the Deep Hedging algorithm, but it can be considered as an alternative problem that best reflects the behavior followed in practice by hedgers, as well as an approximation of the global problem. } Furthermore, the use of a backward approach also allows for the valuation of American options. With respect to American options, it is important to note that the approach proposed by Becker et al. \cite{becker2020pricing} differs from ours, as they rely on the exercise strategy derived from the continuous-time model, rather than adopting the worst-case strategy for the hedger, as we suggest in our proposal.
	
	{Our investigation} includes numerical tests in the one-dimensional Heston model, primarily considered in \cite{buehler2019deep}. The results obtained are consistent with those deduced from the Deep Hedging algorithm and are competitive in terms of quality, confirming the validity of our proposed procedure. {Specifically, {although the solution of the backward problem is usually not optimal for the global  problem,}  numerical results show that the  hedging strategy returned by our algorithm is effective and sometimes even a better minimizer for the global problem than the one obtained with the Deep Hedging algorithm.} It is worth noting that Buehler et al. \cite{buehler2019deep} demonstrate the applicability of the Deep Hedging algorithm in the context of a multidimensional Heston model. Our proposed algorithm, however, cannot efficiently address this case due to the curse of dimensionality, which can only be solved by machine learning-based algorithms. Nevertheless, we believe that our approach may be useful for practitioners when considering low-dimensional problems.

	The paper is organized as follows.  Section \ref{sec:DH}  introduces the pricing framework of the Deep Hedging algorithm. Section \ref{sec:alg} discuss the proposed new algorithm. Section \ref{sec:experiments} is devoted to the experimental
	results. Finally, the conclusions follow in Section
	\ref{sec:conclusions}.
	
	\section{Deep Hedging framework}
	\label{sec:DH}
	
	The hedging model presented in this manuscript is developed within the framework discussed in the Buehler et al. paper \cite{buehler2019deep}. The Deep Hedging algorithm is also used as a point of comparison for our algorithm; therefore, we begin our discussion by briefly presenting it. Although the algorithm is highly adaptable to different stochastic models, we present it with reference to Heston's (one-dimensional) stochastic model, which is also studied in the original manuscript. It is worth noting that in the original version of the algorithm, the authors assume both the drift rate and the interest rate to be zero. Here, we present an updated version of the Deep Hedging algorithm, which works for non-zero drift and interest rate.
	
	\medskip
	The Deep Hedging algorithm allows one to approximate optimal hedging strategies for managing a short position on a European option. The search for an optimal strategy is carried out by exploiting neural networks and it is based on the minimization of a certain loss function. The optimal strategy ultimately determines the option price as the amount of money required to make such a strategy acceptable according to a risk profile. This model makes it possible to determine the value of an option even when perfect hedging is not possible, also when trading costs are considered.

	\subsection{Settings}
	
	We consider a financial market with finite time horizon
	$T$ and trading dates $0=t_{0}<t_{1}<\ldots<t_{N}=T$, with $t_{n}=n\cdot\Delta t$ for $n=0,\dots,N$.
	Let $S^1$ be the stochastic process representing the price of a stock and $V$ the associated variance process. These two stochastic processes are supposed to evolve, under a real-world probability $\mathbb{P}$, according to the Heston model, that is the following stochastic differential equations hold:
	
	\[
	\begin{cases}
		S_{t}^{1}=\mu S_{t}^{1} dt+\sqrt{V_{t}}S_{t}^{1}\mathrm{~d}B_{t}, & S_{0}^{1}=s_0,\\
		V_{t}=\alpha\left(b-V_{t}\right)\mathrm{d}t+\sigma\sqrt{V_{t}}\mathrm{~d}W_{t}, & V_{0}=v_{0},
	\end{cases}
	\]
	where $B$ and $W$ are one-dimensional Brownian motions with their correlation $\rho_{BW}$ a real constant, just as  $\mu, \alpha, b, \sigma, s_0$ {and} $v_0$. In the original version of the manuscript  \cite{buehler2019deep},
	the authors consider a market with a zero drift and interest rate, that is $\mu=r=0$,
	however, here, following R{\'e}millard  and Rubenthaler \cite{remillard2009optimal}, we allow these values to be non-null. Unlike process $S^1$, process $V$ is not directly tradable, {nonetheless} we assume that an idealized variance swap is available in the market, and we denote its price by $S^2$.
	In particular, the maturity of the variance swap is the same as the considered derivative, i.e.  $T$, and if we define the filtration $ \mathcal{F}_t:=\sigma\left(\left(S_s^1, V_s\right): s \in[0, t]\right)$, then
	$$
	S_t^2:= {e^{-r(T-t)}}\cdot\mathbb{E}_{\mathbb{Q}}\left[\int_0^T V_s \mathrm{~d} s \mid \mathcal{F}_t\right], \quad t \in[0, T].
	$$
{	with $\mathbb{Q}$ a risk-neutral probability}.
	Moreover, {if we assume the market price of risk for the variance process to be zero,} one can prove that the idealized price of the variance swap writes down
	$$
	S_t^2= {e^{-r(T-t)}}\cdot\left[ \int_0^t V_s \mathrm{~d} s+L\left(t, V_t\right)\right] 
	$$
	where the function $L$ is defined by
	$$
	L(t, v)=\frac{v-b}{\alpha}\left(1-\mathrm{e}^{-\alpha(T-t)}\right)+b(T-t) .
	$$
	Then,  $S^1$ and $ S^2$ are the prices of tradeable assets that can be used to set up  a {portfolio to hedge a short position on a derivative}. 
	
	Now, let us consider a European option, whose payoff is a random
	variable $\mathcal{F}_{T}$ measurable, denoted by  $\varphi\left(S^1_T \right) $. We assume the position of
	an economic agent who sold the option, so $\varphi\left(S^1_T \right)$ is a liability, and we have to hedge such a short position.
	
	\subsection{Trading strategies}
	
	In order to hedge a liability $\varphi\left(S^1_T \right)$ at $T$, one trades $S^1$ and $S^2$ 
	by using an $\mathbb{R}^{2}$-valued $\mathcal{F}$-adapted stochastic
	process $\delta=\left(\delta_{n}\right)_{n=0,\ldots,N-1}$ with $\delta_{n}=\left(\delta_{n}^{1},\delta_{n}^{2}\right)$.
	Please observe that, $\delta_{n}^{i}$ denotes the agent's holdings
	of the $i$-th asset, after rebalancing the portfolio at time $t_n$. For simplicity of notation,
	we may also define $\delta_{-1}^{i}=\delta_{N}^{i}:=0$ for $i=1,2,$
	which implies an empty portfolio at beginning and full liquidation at maturity.
	
	Let $p_{0}$ denote the cost for building the portfolio {at} $t_{0}$,
	so $p_{0}$ is the amount of money required in payment by the buyer of the option. In a market
	without transaction costs the agent's {global} profit and loss $PL_0^T$ at time $T$
	is thus given by $$PL_0^T\left(\varphi\left(S^1_T \right),p_0,\delta \right) =-\varphi\left(S^1_T \right)+p_{0}e^{rT}+(\delta\cdot S)_{0}^{N},$$ where  
	\[
	(\delta\cdot S)_{n_1}^{n_2}:=\sum_{k=n_1}^{n_2-1} \left[ \delta_{k}^{1}\left(S_{t_{k+1}}^{1}-S_{t_{k}}^{1}e^{r\Delta t}\right)+\delta_{k}^{2}\left(S_{t_{k+1}}^{2}-S_{t_{k}}^{2}e^{r\Delta t}\right)\right] e^{\left(T-t_{k+1}\right)r}
	\]
	is the profit due to the hedging strategy, with respect to the time intervals $\left[t_k,t_{k+1} \right] $ for $k=n_1,\dots,n_2-1 $, capitalised to maturity. 
	
	Trading costs can also be included in the model. We assume that any
	trading activity causes {proportional} costs as follows: if the hedger decides to
	buy or sell  $\alpha\in\mathbb{R}$ units of $S^i$ at time $t_{k}$,
	then he incurs in a transaction cost which amounts to $\varepsilon\cdot S_k^i \cdot \left|\alpha \right|$, with $\varepsilon$ the proportional cost for trading $1$ currency unit of asset. The capitalised to maturity cost of the trading
	strategy $\delta$,  {with respect to} trading times from $t_{n_1}$ to $t_{n_2}$, is {therefore} given by 
	\[
	C_{n_1}^{n_2}\left(\delta\right):=\sum_{k=n_1}^{n_2} \varepsilon \left[  S_{t_k}^{1}\left|\delta_{k}^{1}-\delta_{k-1}^{1}\right|+S_{t_k}^{2}\left|\delta_{k}^{2}-\delta_{{k-1}}^{2}\right| \right] e^{\left(T-t_k\right)r}.
	\]
	We stress out that, according to the original paper, $C_{0}^{N}\left(\delta\right)$  includes the cost of setting up the portfolio at time $0$ and   the costs of divestment at time $T$.
	{As a result}, when transaction costs are applied, the agent's terminal portfolio value at $T$ {reads out}
	\[
	{PL}_{0}^{T}\left(\varphi\left(S^1_T \right),p_{0},\delta\right):=-\varphi\left(S^1_T \right)+p_{0}e^{rT}+(\delta\cdot S)_{0}^{N}-C_{0}^{N}(\delta).
	\]
	Such a random variable can be considered as the error of the trading strategy, and one seeks to minimize it.
	
	\subsection{Hedging and pricing}
	
	The basic situation to considered is given by a complete market with
	continuous-time trading, no transaction costs, and unconstrained hedging.
	In this particular case, for any liability $\varphi\left(S^1_T \right)$, there exists a unique
	replication strategy $\delta$ and a unique fair price $p_{0}\in\mathbb{R}$
	such that $$-\varphi\left(S^1_T \right)+p_{0}e^{rT}+(\delta\cdot S)_{0}^{N}=0$$ holds $\mathbb{P}$-a.s..
	More generally, the optimal hedging strategy $\delta$ and the option price $p_{0}$ can
	be defined as the solution of the following {problem}: computing $p_{0}\in\mathbb{R}$ and $\delta\in\mathcal{H}$ such that 
	\begin{equation}
		\mathbb{E}\left[\left(-\varphi\left(S^1_T \right)+p_{0}e^{rT}+(\delta\cdot S)_{0}^{N}\right)^{2}\right]=0, 
	\end{equation}
	where $\mathcal{H}$ stands for the set of all possible trading strategies.  
	
	In an incomplete market with frictions, perfect hedging is (usually) no more
	possible, so, it is no longer possible to obtain a strategy that generates an almost certainly zero error. Consequently, the new objective is to minimize the error of the hedging strategy, so, one considers the problem of minimizing the mean squared error of the strategy: 
	\begin{equation}
		\inf_{p_{0}\in\mathbb{R}}\inf_{\delta\in\mathcal{H}}\mathbb{E}\left[\left(-\varphi\left(S^1_T \right)+p_{0}e^{rT}+(\delta\cdot S)_{0}^{N}-C_0^N(\delta)\right)^{2}\right].\label{eq:L2}
	\end{equation}
	The solution ${p}_{0}$ of \eqref{eq:L2}  can be regarded as the fair hedging price, but it does not take into account the non-eliminable risk faced by the hedger. In this case,
	one  has to specify a certain criterion that defines an acceptable
	``minimal price'' for any liability, a price that includes compensation
	for the risk assumed by the hedger. Such a price, can be defined as the minimum
	amount of money to add to the hedging portfolio {at the launch of the contract}, so that, {when the optimal strategy is employed}, the overall position becomes acceptable at maturity.
	The criterion for identifying what is an acceptable position is defined by {means of } a convex risk measure $\rho$, such that $\rho\left(0\right)=0$. For example, hereafter, we consider $\rho$ as the $\alpha$-level CVaR, that is $\rho\left(X\right)=\mathrm{CVaR_{\alpha}}\left(X\right)$.
	The problem to solve is then {computing}
	\begin{equation}
		\inf_{\delta\in\mathcal{H}}\rho\left(-\varphi\left(S^1_T \right)+(\delta\cdot S)_{0}^{N}-C_{0}^{N}(\delta)\right),\label{eq:RM}
	\end{equation}
	{so that} the hedger's price writes down
	\begin{equation}
		\hat{p}_{0}=e^{-rT}\inf_{\delta\in\mathcal{H}}\rho\left(-\varphi\left(S^1_T \right)+(\delta\cdot S)_{0}^{N}-C_{0}^{N}(\delta)\right).\label{eq:RMP}
	\end{equation}
	
	Let $\hat{\delta}\in\mathcal{H}$ be the optimal strategy, solution of problem \eqref{eq:L2} or \eqref{eq:RM}, according to the evaluation approach under investigation.
	Please, observe that the optimal hedging strategy $\hat{\delta}_{n}^{i}$ at time $t_{n}$
	can be seen as the outputs of a function that depends on filtration
	at time $t_{n}$: $\hat{\delta}_{n}^{i}=\hat{\delta}_{n}^{i}\left(S^1_{0},S^2_{0},\dots,S^1_{n},S^2_{n}\right)$. Accordingly, problems (\ref{eq:L2}) and (\ref{eq:RM}) can be interpreted
	as variational problems, which can be tackled by
	exploiting a reinforcement learning algorithm based on neural networks.
	
	\subsection{The Deep Hedging algorithm}
	
	Neural  networks can be used
	to solve variational problems such as (\ref{eq:L2}) and (\ref{eq:RM})
	by choosing a suitable loss function. 
	
	\smallskip
	The first step consists in simulating the paths of the underlyings $\left(S^1,S^2\right)$
	at discrete times $t_{n}$. 
	These simulations are used to train a semi-recurrent neural  network: at each time-step $t_n$, for $n=0,\dots,N-1$ it
	takes as inputs the 4 values of $S_{t_n}^{1}$, $S_{t_n}^{2}$, $\delta_{n-1}^{1}$
	and $\delta_{n-1}^{2}$ and it outputs $ {\delta}_{n}^{1}$ and $ {\delta}_{n}^{2}$.
	{The network is trained to minimize the loss function,  which value is determined based on  the simulations of the paths. At the end of this process, the network returns a hedging strategy, denoted by $\hat{\delta}$, that approximates the solution to the problems (\ref{eq:L2}) or (\ref{eq:RM}).} 
	{Specifically,} the weight matrices and biases of the neurons are the parameters of the neural network but also the initial values $\delta_{0}^{1}$ and
	$\delta_{0}^{2}$ are added among the parameters to be optimize. Now, let $\theta$ be the vector all the parameters of the neural network. So, $\theta$ is computed as a minimum point of a suitable loss function $\ell\left(\theta\right)$. {While} problem (\ref{eq:RM}) is considered, the loss function is $\ell\left(\theta\right)=\rho\left(-\varphi\left(S^1_T \right)+(\delta\cdot S)_{0}^{N}-C_{0}^{N}(\delta)\right)$. As far as problem (\ref{eq:L2}) is considered, also $p_{0}$ is included
	among the parameters to be optimized and  the loss function is set
	as $\ell\left(\theta\right)=\mathbb{E}\left[\left(-\varphi\left(S^1_T \right)+p_{0}e^{rT}+(\delta\cdot S)_{0}^N-C_{0}^{N}\right)^{2}\right]$.  
	The neural network is then trained by using the ADAM algorithm (Kingma and Ba \cite{kingma2015adam}). 
	Finally, when considering problem (\ref{eq:RM}), the trained neural network is tested on out-of-sample dataset to determine the price $p_0$, so that the measurement of network performance is not altered by overfitting.
	
	\medskip
	Once the neural network has been trained, it returns the following outputs: the
	option prince $p_{0}$ and the initial values $\hat{\delta}_{0}^{1}$ and
	$\hat{\delta}_{0}^{2}$ as well as it can be employed to  calculate delta values at each possible trading time-step. This means that the neural network can be used for hedging purposes:
	given a path $\left(S_{t_n}^{1},S_{t_n}^{2}\right)_{n=0,\ldots,N}$ or
	even a partial path $\left(S_{t_k}^{1},S_{t_k}^{2}\right)_{k=0,\ldots,n}$
	(with $n\leq N$), the neural network returns its optimal values for $\left(\hat{\delta}_{k}^{1},\hat{\delta}_{k}^{2}\right)_{k=0,\ldots,n}$,
	so one can effectively exploit the Deep Hedging algorithm for hedging a short position.

	\section{Backward Hedging}
	\label{sec:alg}
	The Deep Hedging algorithm relies on  neural networks to achieve global optimisation of the hedging strategy. This means that the algorithm simultaneously optimises all the parameters {$\theta$} that define the entire hedging strategy. This approach results in a single high-dimensional optimisation problem.
	
	The algorithm we propose, on the other hand, solves a large number of optimisation problems in only two variables, that is $\left(\delta_{n}^{1},\delta_{n}^{2}\right)$. Starting from a specific configuration of the market state, it determines the optimal delta values for that specific configuration. Thanks to linear interpolation,  the hedging strategy is extended to the entire space of interest. 
	We emphasise that, unlike other works already present in the literature (see e.g. Potter et al. \cite{potters2001hedged}) that consider the loss between two trading times, the loss function to be optimised is {evaluated at maturity}, i.e. it considers the entire hedging strategy, from the considered time-step until maturity. 
	
	Finally, we would like to point out that the pseudocodes of the main versions of the Backward Hedging algorithm, discussed below, are included in the concluding Appendix \ref{sec:pc}.
	
	\subsection{The backward approach}\label{ba}
	 We begin our discussion by presenting the proposed hedging model for the special case $\varepsilon=0$, that is no transaction costs {and loss function based on the risk measure $\rho$, as in problem \eqref{eq:RM}}.

	 {The hedging strategy we propose is determined by proceeding backward in time, starting from time $t_{N-1}$, to the initial instant t=0. At the generic hedging instant $t_n$, the amount of hedging instruments to be held in the portfolio $\delta^1_n$ and $\delta^2_n$ are determined by minimizing an appropriate measure of risk relative to the future hedging error, that is, between $t_n$ and $T$. Specifically, the Backward Hedging strategy $\bar{\delta}$ at time $t_n$ is defined as the solution of the following problem}:
\begin{multline}
	\left(\bar{\delta}_{n}^{1},\bar{\delta}_{n}^{2}\right)=	\argmin_{\left(x,y\right)\in \mathbb{R}^2}\rho\left(-\varphi\left(S^1_T \right)+(\bar{\delta}\cdot S)_{n+1}^{N}+\right. 	\left.\left[ x\left(S_{t_{n+1}}^{1}-S_{t_{n}}^{1}e^{r\Delta t}\right)+
	y\left(S_{t_{n+1}}^{2}-S_{t_{n}}^{2}e^{r\Delta t}\right)\right] e^{\left(T-t_{n+1}\right)r} \right). \label{eq:nN}
\end{multline} 
 
	{Let us remark that in equation  \eqref{eq:nN}, at time $t_n$, $S_{t_{n}}^{1}$ and $S_{t_{n}}^{2}$ are measurable, while $S_{t_{k}}^{1}$ and $S_{t_{k}}^{2}$ for $k>n$ are random variables.} 
	We, observe that problem \eqref{eq:nN} does not involve the previous hedging strategy, but only the future hedging strategy, i.e. $\left(\bar{\delta}_{k}^{1},\bar{\delta}_{k}^{2}\right)$ for $k= n+1,\dots,N$. { We stress out that excluding past history seems reasonable from a practical perspective since an hedger  decides his strategy looking only at future results. In fact, the gain and the losses already realised through the hedging strategy are measurable values and, due to the cash invariance property of risk measures, their inclusion/exclusion does not change the solution to  problem \eqref{eq:nN}. Moreover,  problem \eqref{eq:nN} is suitable to be tackled through a backward procedure: if one solves this kind of problems backward in time, future hedging strategy should be  available as referred to time steps already treated.}  
	Furthermore, the {applicability} of a backward approach is also ensured by the fact that Heston's model is Markovian, so the future evolution of processes, conditioned by the state of the present market, is independent of the past. We can therefore observe that values $\left(\bar{\delta}_{k}^{1},\bar{\delta}_{k}^{2}\right)$  only depend on the current state of the market.	
	{Once one has solved problem \eqref{eq:nN} for $t_n=t_N,\dots,0$, the Backward Hedging option price can be computed as 
		\begin{equation}
		\bar{p}_{0}=e^{-rT} \rho\left(-\varphi\left(S^1_T \right)+(\bar{\delta}\cdot S)_{0}^{N}-C_{0}^{N}(\bar{\delta})\right).\label{eq:BHP}
\end{equation}}
	
	\medskip
	{Before presenting the resolution of the problem \eqref{eq:nN} in Heston's model, we discuss the relation between the strategies $\hat{\delta}$ and $\bar{\delta}$, that is the solution of the global hedging problem \eqref{eq:RM} and of the backward problem  \eqref{eq:nN},  respectively. We now show that the two strategies, in general, are not equivalent. }
		
		\subsection{A key example} {We consider, as an example, a discrete time model, with only 3 trading times, namely $t_0=0, t_1=1$ and $t_2=2$. For this example only, the market includes a single underlying with initial value $S_0=1$. The underlying can assume $5$ equiprobable values at time $t_1$, which are  $S_1^i=S_0\cdot u^{2i-6}$, for $i=1,\dots,5$ and $u=\exp\left(0.2\sqrt{1/4} \right)$. Moreover, for each value of $S_1^i$, there are $5$ equiprobable continuation values, namely $S_2^{i,j}=S_1^i\cdot u^{2j-6}$, for $j=1,\dots,5$. So there are $5$ possible paths from   $t_0$ to $t_1$ and $25$ possible paths from   $t_0$ to $t_2$.    We stress  out that these values are the same as those generated at time-steps $0,4$ and $8$ by the CRR model with $T=2$ and $N=8$ discretization steps, with the only difference being that, to simplify the calculations, we assume transaction probabilities between two consecutive instants all equal to $1/5$. Also, for simplicity, we assume the interest rate $r$ to be zero. \\
    In this model, an hedging strategy consists of $6$ values: $\delta_0$, i.e. the amount of stocks to include in the hedging portfolio at time $t=0$, and $\delta_{1,i}=\delta_1(S_1^i), i=1,\dots,5$, i.e. the  amount of stocks to include in the hedging portfolio at time $t_1$ according to the $5$ possible values of the underlying at that time.}   
       
    { Now, we consider a short position on a strangle, i.e. a short position on a call and on a put, with strikes  $K_C=0.8$ and $K_P=1.3$ respectively. 
     In our example, we   want to hedge the strangle so as to minimize the $40\%$-CVaR. For the global-optimal strategy  $\hat{\delta}$, this consists of choosing the $6$ hedging parameters so as to minimize the CVaR relative to profit-and-loss between $t_0$ and $t_2$, which we denote by $\mathrm{CVaR}_0^2$. For the backward strategy $\bar{\delta}$, on the other hand, this consists of determining at time $t_1$, for each of the possible values of the underlying $S_1^i$, the hedging parameter $\bar{\delta}_1(S_1^i)$ so as to minimize the CVaR relative to the profit-and-loss between $t_1$ and $t_2$, which we denote by $\mathrm{CVaR}_1^2$, and then the hedging parameter $\bar{\delta}_0$ at time $t_0$ so as to minimize the $\mathrm{CVaR}_0^2$ between $t_0$ and $t_2$, being subject to the constraint of using  $\bar{\delta}_1(S_1^i)$  as the hedging strategy at time $t_1$. {Let us point that}, since we are assuming equal transaction probabilities and $40\%$ as CVaR level, then $\mathrm{CVaR}_0^2$ is equal to the opposite of the average of the 10 lower profit-and-loss outcomes between $t_0$ and $t_2$, while $\mathrm{CVaR}_1^2$ is equal to the opposite of the average of the 2 lower profit-and-loss outcomes between $t_1$ and $t_2$.\\       
     The two optimal strategies are shown in Table \ref{tab:hs},  while Table \ref{tab:comp} reports the computations for those strategies. Specifically, the first three columns show the 5 possible values of $S_1$, the 25 final values of $S_2$ and their payoffs, respectively. The next columns report the profit and loss values for each of the $25$ possible trajectories, calculated by applying the 2 strategies under consideration. These profit-and-loss are calculated  between $t_0$ and $t_2$, denoted by $PL_0^2$, or between $t_1$ and $t_2$, denoted by $PL_1^2$. Values in bold are the worst values, and they contribute to the calculation of CVaR values.  For both the two strategies, $\mathrm{CVaR}_1^2$ is reported for each of the $5$ possible values of $S_1$, while the two values for $\mathrm{CVaR}_0^2$  are reported at the bottom of the Table. \\
     We can observe that the two strategies are successful in their respective problems: $\hat{\delta}$ generates a smaller $\mathrm{CVaR}_0^2$ than $\bar{\delta}$, while $\bar{\delta}$ generates smaller (or equal) $\mathrm{CVaR}_1^2$ for all possible values of $S_1$. We can therefore conclude that the two strategies $\hat{\delta}$ and $\bar{\delta}$, and therefore the problems \eqref{eq:RM} and \eqref{eq:nN}, are not equivalent.}  
 
	\begin{table}\begin{center} 
	\begin{tabular}{ccc}
		\toprule 
		Index & $\hat{\delta}$ & $\bar{\delta}$\tabularnewline
		\midrule
		$0$ & 0.1468 & 0.1468\tabularnewline
		$1,1$ & 0.9217 & 0.9217\tabularnewline
		$1,2$ & 0.5203 & 0.6351\tabularnewline
		$1,3$ & 0.0650 & 0.2850\tabularnewline
		$1,4$ & -0.3733 & -0.2353\tabularnewline
		$1,5$ & -0.8671 & -0.5567\tabularnewline
		\bottomrule
\end{tabular}\end{center}
\caption{\label{tab:hs} The hedging strategies $\hat{\delta}$ and $\bar{\delta}$.}
\end{table}

	\begin{table}\begin{center} 
		\begin{tabular}{lllllllll}
			\toprule 
			&  &  &  & \multicolumn{2}{l}{Strategy $\hat{\delta}$}  &  & \multicolumn{2}{l}{Strategy $\bar{\delta}$} \tabularnewline
			\midrule
			$S_{1}$ & $S_{2}$ & Payoff &  & $PL_{0}^{2}$ & $PL_{1}^{2}$ &  & $PL_{0}^{2}$ & $PL_{1}^{2}$\tabularnewline
			\midrule
			& 2.2255 & 1.4255 &  & \textbf{-0.6771} & -0.7493 &  & -0.6771 & -0.7493\tabularnewline
			& 1.8221 & 1.0221 &  & -0.6455 & -0.7177 &  & -0.6455 & -0.7177\tabularnewline
			1.4918 & 1.4918 & 0.6918 &  & -0.6196 & -0.6918 &  & -0.6196 & -0.6918\tabularnewline
			& 1.2214 & 0.5000 &  & \textbf{-0.6771} & \textbf{-0.7493} &  & \textbf{-0.6771} & \textbf{-0.7493}\tabularnewline
			& 1.0000 & 0.5000 &  & \textbf{-0.8811} & \textbf{-0.9533} &  & \textbf{-0.8811} & \textbf{-0.9533}\tabularnewline
			\cmidrule{2-9} \cmidrule{3-9} \cmidrule{4-9} \cmidrule{5-9} \cmidrule{6-9} \cmidrule{7-9} \cmidrule{8-9} \cmidrule{9-9} 
			&  &  & $\mathrm{CVaR}_{1}^{2}$ &  & 0.8513 &  &  & 0.8513\tabularnewline
			\midrule
			& 1.8221 & 1.0221 &  & \textbf{-0.6770} & \textbf{-0.7095} &  & -0.6081 & -0.6406\tabularnewline
			& 1.4918 & 0.6918 &  & -0.5186 & -0.5511 &  & -0.4876 & -0.5201\tabularnewline
			1.2214 & 1.2214 & 0.5000 &  & -0.4675 & -0.5000 &  & -0.4675 & -0.5000\tabularnewline
			& 1.0000 & 0.5000 &  & -0.5827 & -0.6152 &  & -0.6081 & \textbf{-0.6406}\tabularnewline
			& 0.8187 & 0.5000 &  & -0.6770 & \textbf{-0.7095} &  & \textbf{-0.7232} & \textbf{-0.7557}\tabularnewline
			\cmidrule{2-9} \cmidrule{3-9} \cmidrule{4-9} \cmidrule{5-9} \cmidrule{6-9} \cmidrule{7-9} \cmidrule{8-9} \cmidrule{9-9} 
			&  &  & $\mathrm{CVaR}_{1}^{2}$ &  & 0.7095 &  &  & 0.6982\tabularnewline
			\midrule
			& 1.4918 & 0.6918 &  & -0.6599 & \textbf{-0.6599} &  & -0.5517 & -0.5517\tabularnewline
			& 1.2214 & 0.5000 &  & -0.4856 & -0.4856 &  & -0.4369 & -0.4369\tabularnewline
			1.0000 & 1.0000 & 0.5000 &  & -0.5000 & -0.5000 &  & -0.5000 & -0.5000\tabularnewline
			& 0.8187 & 0.5000 &  & -0.5118 & -0.5118 &  & -0.5517 & \textbf{-0.5517}\tabularnewline
			& 0.6703 & 0.6297 &  & -0.6511 & \textbf{-0.6511} &  & \textbf{-0.7236} & \textbf{-0.7236}\tabularnewline
			\cmidrule{2-9} \cmidrule{3-9} \cmidrule{4-9} \cmidrule{5-9} \cmidrule{6-9} \cmidrule{7-9} \cmidrule{8-9} \cmidrule{9-9} 
			&  &  & $\mathrm{CVaR}_{1}^{2}$ &  & 0.6555 &  &  & 0.6376\tabularnewline
			\midrule
			& 1.2214 & 0.5000 &  & -0.6769 & \textbf{-0.6503} &  & -0.6214 & -0.5948\tabularnewline
			& 1.0000 & 0.5000 &  & -0.5943 & -0.5677 &  & -0.5693 & -0.5427\tabularnewline
			0.8187 & 0.8187 & 0.5000 &  & -0.5266 & -0.5000 &  & -0.5266 & -0.5000\tabularnewline
			& 0.6703 & 0.6297 &  & -0.6009 & -0.5743 &  & -0.6214 & \textbf{-0.5948}\tabularnewline
			& 0.5488 & 0.7512 &  & \textbf{-0.6770} & \textbf{-0.6504} &  & \textbf{-0.7143} & \textbf{-0.6877}\tabularnewline
			\cmidrule{2-9} \cmidrule{3-9} \cmidrule{4-9} \cmidrule{5-9} \cmidrule{6-9} \cmidrule{7-9} \cmidrule{8-9} \cmidrule{9-9} 
			&  &  & $\mathrm{CVaR}_{1}^{2}$ &  & 0.6504 &  &  & 0.6412\tabularnewline
			\midrule
			& 1.0000 & 0.5000 &  & \textbf{-0.8343} & \textbf{-0.7859} &  & \textbf{-0.7319} & \textbf{-0.6835}\tabularnewline
			& 0.8187 & 0.5000 &  & \textbf{-0.6771} & -0.6287 &  & -0.6310 & -0.5826\tabularnewline
			0.6703 & 0.6703 & 0.6297 &  & \textbf{-0.6781} & -0.6297 &  & \textbf{-0.6781} & -0.6297\tabularnewline
			& 0.5488 & 0.7512 &  & \textbf{-0.6942} & -0.6458 &  & \textbf{-0.7319} & -0.6835\tabularnewline
			& 0.4493 & 0.8507 &  & \textbf{-0.7074} & \textbf{-0.6590} &  & \textbf{-0.7760} & \textbf{-0.7276}\tabularnewline
			\cmidrule{2-9} \cmidrule{3-9} \cmidrule{4-9} \cmidrule{5-9} \cmidrule{6-9} \cmidrule{7-9} \cmidrule{8-9} \cmidrule{9-9} 
			&  &  & $\mathrm{CVaR}_{1}^{2}$ &  & 0.7225 &  &  & 0.7056\tabularnewline
			\midrule
			&  &  & $\mathrm{CVaR}_{0}^{2}$  & \multirow{1}{*}{0.7180} &  &  & \multirow{1}{*}{0.7314} & \tabularnewline
			\bottomrule
		\end{tabular} \end{center}
	\caption{\label{tab:comp}Comparison between hedging strategies $\hat{\delta}$ and $\bar{\delta}$.}
	\end{table}
	\medskip
	
	\subsection{Comparison between strategies}
	
	{At this point, it is worth asking which of the two strategies is the most correct to consider for practical purposes. The global optimal strategy, pursued by the Deep-Hedging algorithm, hides a pitfall.\\		
		The strategy $\hat{\delta}$ is optimal at time zero, that is between $t_0=0$ and $T$, but only at time zero and has no optimality property at times $t_n>0$, between $t_n>0$ and $T$: in fact, as shown by the previous example, the   strategy $\hat{\delta}$ is not optimal at time $t_1$, as it does not always minimize $\mathrm{CVaR}_{1}^{2}$, in as exceeded by $\bar{\delta}$. If we assume that, at any time $t_n> 0$, the hedger wishes to minimize the CVaR between $t_n$ and $T$, then he will {be led to abandon strategy $\hat{\delta}$ that}  he had decided at time $0$, in favor of a more performing one. The initial valuation is therefore not consistent with what is expected of the hedger: the option is priced based on a strategy which will not be followed afterwards. This annoying phenomenon is related to the absence of time-consistency for CVaR, and more generally for the most common risk measures.\\
		The $\bar{\delta}$ strategy, on the other hand, at time $t_n$ is optimal given the strategy that will be followed subsequently, therefore it is not convenient to modify it, unlike $\hat{\delta}$, and it is less penalizing in future times. Once established, it should be maintained (unless it is replaced with another $\hat{\delta}$ strategy recomputed to be optimal between $t_n$ and $T$, which however will be disregarded in subsequent times $t_k>t_n$). This aspect could be appreciated by the supervisory bodies which must validate the hedging procedure.  \\ Finally, we point out that Basak and Chabakauri \cite{basak2010dynamic} and subsequently Cui et al. \cite{cui2019time}, have come to similar considerations in the context of portfolio allocation: although the global-optimal strategy first appears to have strong global investment performance, investors would deviate from such a strategy throughout the investment process.\\
	For the reasons previously mentioned, we propose to use $\bar{\delta}$ in place of  $\hat{\delta}$ .}
	
	\subsection{European option with no transaction costs, based on risk measure}\label{ss1}
	{We discuss the numerical procedure we propose to solve problem  \eqref{eq:nN}.	}
	The issues that one has to face are controlling the hedging strategy for future time-steps, computing the objective function, that is the risk measure, and optimizing the objective function.
	
	First of all, we propose to estimate the risk measure $\rho$ through (forward) $M$ Monte Carlo simulations, from $t_n$ to $T$, of the underlying processes $S^1$ and $S^2$ conditioned by $\mathcal{F}_{t_n}$. In order to simulate future values of $S^1,S^2$, one only needs to know $S^1_{t_n},V_{t_n}$ and $I_{t_n}=\int_{0}^{t_n}V_s \mathrm{d}s$ in place of all filtration $\mathcal{F}_{t_n}$, so  the market state can be described by this triple. Please, note that knowing $\left(S^1_{t_n},S^2_{t_n}\right)$ is not sufficient to simulate future values $\left\lbrace \left(S^1_{t_k},S^2_{t_k}\right)|\mathcal{F}_{t_n}, k=n+1,\dots,N\right\rbrace $, as, to simulate the underlying $S^1_{t_k}$, one must first simulate the volatility $V_{t_k}$, and the initial value of the volatility $V_{t_n}$ cannot be deduced from $\left(S^1_{t_n},S^2_{t_n}\right)$.
	Therefore we also generate the values of $I$ and $V$, and denote the simulated values with  $$\mathcal{S}_n=\left\lbrace  \left(S^1_{k,m}, S^2_{k,m}, I_{k,m},V_{k,m} \right), k=n+1,\dots,N, \ m=1,\dots,M \right\rbrace.$$ Now, for each time-step $t_n$, we consider a grid of values $\mathcal{G}_n=\mathcal{G}_n^S\times \mathcal{G}^I_n\times \mathcal{G}^V_n$ which is defined as follows: $\mathcal{G}_n^S$ is a uniform mesh of $N_S$ points for $S^1_{t_n}$, $\mathcal{G}_n^I$ is a uniform mesh of $N_I$ points for $I_{t_n}$ and $\mathcal{G}_n^V$ is a mesh of $N_V$ points for $V_{t_n}$.  The maximum and minimum values of $\mathcal{G}_n^S$ and $\mathcal{G}^I_n$ are set by computing appropriate quantiles (in our tests we consider $0.1\%$ and $99.9\%$ quantiles) for  $S^1_{t_n}$ and  $I_{t_n}$. Those quantiles are estimated from $M_0$  Monte Carlo  simulations of the Heston model, starting from the initial values $S_0^1=s_0$ and $V_0=v_0$, which we denote with $$\mathcal{\hat{S}}_n=\left\lbrace  \left(\hat{S}^1_{k,m}, \hat{I}_{k,m} \right), k=1,\dots,N, \ m=1,\dots,M_0 \right\rbrace.$$ Simulations of the stochastic processes are obtained through the hybrid Monte Carlo-Tree method introduced by Briani et al. \cite{briani2017hybridapproach,briani2017hybridtree}, together with the trinomial tree  introduced by Goudenège et al. \cite{goudenege2021gaussian}, by using $N_T$ simulation time-steps ($N_T$ must be a multiple of $N$, so that the simulated values at the trading epochs can be obtained by subsampling). This simulation method is particularly interesting in that the possible states for the random variable $V_{t_n}$ are discretized through the use of a trinomial tree $\mathcal{T}$ that exactly reproduces the first and second moments of the transition distribution of the process $V$, thus leading to accurate results with only a few time-discretization steps. This trinomial tree is also used to define the grid $\mathcal{G}^V_n$ as the values of $V$ associated to the tree nodes at time $t_n$. Moreover, it is worth observing  that the simulated values of $V$ always belong to the grid. Finally, we also set $\mathcal{G}_0=\left\lbrace s_0 \right\rbrace \times \left\lbrace 0\right\rbrace \times \left\lbrace v_0 \right\rbrace$ for the market state at inception.
	
	Problem  \eqref{eq:nN} is solved for each point of $\mathcal{G}$ backward in time so, at time $t_{n}$, the hedging strategy is known at future time-steps $\left\lbrace t_{k}\right\rbrace _{k=n+1,\dots,N-1}$, but only for those values of $\left(S^1_{t_k},I_{t_k},V_{t_k} \right) $ that are points of $\mathcal{G}_k$. However, the simulated values of $S_{t_k}$ and $I_{t_k}$   for the estimation of the risk measure may  not belong to the grid $\mathcal{G}_k$ and therefore the hedging strategy would not be known for those special values. This problem is overcome by using linear interpolation, which allows one to extend the hedging strategy from $\mathcal{G}_k$ to all $\mathbb{R}^3$. In this regard, we point out that it is necessary to resort to an interpolation in the $S$ and $I$ coordinates only, as the simulated values of $V$ belong to the grid thanks to the use of the trinomial tree $\mathcal{T}$ for simulation.
	We also point out that although other forms of interpolation are possible (see, e.g., Akima \cite{akima1974method, akima1978method} for Modified Akima interpolation or Press et al. \cite{press2007numerical} for Bicubic Interpolation), linear interpolation is chosen for its low computational cost. This element is decisive as the proposed algorithm requires the calculation of a large number of interpolations.  
	As a result, we can represent the hedging strategy as a collection of function $\left\lbrace \left(\bar{\delta}^1_n,\bar{\delta}_n^2 \right):\mathbb{R}^3\rightarrow\mathbb{R}^2 \right\rbrace $ such that $\left(\bar{\delta}^1_n,\bar{\delta}_n^2 \right)\left(s,i,v \right) $ are the {optimal hedging values}  when $S^1_{t_n}=s,I_{t_n}=i$ and $V_{t_n}=v$.
	
	Now that we have clarified how to calculate the objective function, we must solve the optimization problem \eqref{eq:nN}, thus determining the specific values $\left(\bar{\delta}^1_n,\bar{\delta}_n^2 \right)$ for the grid point under evaluation. This is done by exploiting a standard numerical procedure. Specifically, we used the Simplex Search method of Lagarias et al. \cite{lagarias1998convergence}, which is an efficient derivative--free algorithm and it is particularly suitable for problems of this kind.
	
	Summing up, the Backward Hedging algorithm proceeds backward in time by solving problems  \eqref{eq:nN}  for all nodes in the grid, starting from time $t_{N-1}$ down to time $t_0$. Once the hedging strategy has been determined for all trading times, we simulate a new sample of $M_1$  Monte Carlo simulations 
	$$\mathcal{\bar{S}}_0=\left\lbrace  \left(\bar{S}^1_{k,m}, \bar{S}^2_{k,m}, \bar{I}_{k,m}, \bar{V}_{k,m} \right), k=1,\dots,N, \ m=1,\dots,M_1 \right\rbrace $$
	and we apply the algorithm on this new out-of-sample data. The risk measure determined on this new sample defines the initial option price {$\bar{p}_0$ with respect to the Backward Hedging algorithm}.  
	
	\subsection{Loss as mean squared error}
	The Backward Hedging algorithm can also be employed to {tackle} problem \eqref{eq:L2}, {by a backward approach}. Again, let us again assume that no transaction costs are applied. It is easy to prove that the optimal value of $p_0$ according to \eqref{eq:L2} is given by 
	\begin{equation}
		p_0=-e^{-rT}\mathbb{E}\left[-\varphi\left(S^1_T \right)+(\delta\cdot S)_{0}^{N} \right],
	\end{equation}
	and for this choice of $p_0$, we obtain
	\begin{equation*}
		\mathbb{E}\left[\left(-\varphi\left(S^1_T \right)+p_{0}e^{rT}+(\delta\cdot S)_{0}^{N}\right)^{2}\right]=\mathbb{V}\mathrm{ar}\left[-\varphi\left(S^1_T \right)+(\delta\cdot S)_{0}^{N}\right],
	\end{equation*}
	thus \eqref{eq:L2} reduces to solving
	\begin{equation}
		\inf_{\delta\in\mathcal{H}}\mathbb{V}\mathrm{ar}\left[ -\varphi\left(S^1_T \right)+(\delta\cdot S)_{0}^{N}\right].\label{eq:var}
	\end{equation}
{Now, let us consider the backward version of problem \eqref{eq:var}, that is at time $t_n$ one has to solve the following problem}	\begin{multline}
	\left(\bar{\delta}_{n}^{1},\bar{\delta}_{n}^{2}\right)=	\argmin_{\left(x,y\right)\in \mathbb{R}^2}\mathbb{V}\mathrm{ar}\left[-\varphi\left(S^1_T \right) +(\bar{\delta}\cdot S)_{n+1}^{N}+\right.  	\left.\left[ x\left(S_{t_{n+1}}^{1}-S_{t_{n}}^{1}e^{r\Delta t}\right)+
	y\left(S_{t_{n+1}}^{2}-S_{t_{n}}^{2}e^{r\Delta t}\right)\right] e^{\left(T-t_{n+1}\right)r}\right].\label{eq:var2}
\end{multline} 
	{Problem \eqref{eq:var2} is of the same type of \eqref{eq:nN}, with the only difference being that  $\rho$ is replaced with $\mathbb{V}\mathrm{ar}$, so one can tackle it   again with the Backward Hedging algorithm, as described in the previous Subsection}.
	
	\subsection{American options}
	American options can be exercised at any moment before maturity. With the scope of pricing, we approximate such a feature with a weaker one:  we suppose that the option can be exercised at any trading time. If the option is exercised, the hedger pays the payoff to the buyer, otherwise he rebalances the hedging portfolio. 
	Pricing requires to model the exercising strategy, which depends on the buyer. Here, we assume the so called \textit{worst-hedging-case}, which means that the option is exercised in the worst case for the hedger, that is as soon as the payoff is higher than the hedging cost, which corresponds to the continuation cost.
	Specifically, the option price  $\bar{p}_n$ at time $t_n$ is define as a function of $S^1_{t_n},I_{t_n},V_{t_n}$, {again as the cost for hedging the position, {from $t_n$ up the the exercise time $\tau_n$}} that is
	\begin{equation} 
		\bar{p}_n\left(S^1_{t_n},I_{t_n},V_{t_n} \right)= 
		\inf_{\delta\in\mathcal{H}}\rho\left(\left( -\varphi\left(S^1_{\tau_n} \right)+(\bar{\delta}\cdot S)_{0}^{\tau_n/\Delta t}\right)e^{-r \left( \tau_n-t_n\right) } \right).
	\end{equation}
	{Specifically}, $\tau_n$  is the stopping time defined as
	\begin{equation} 
		\tau_n=\min\left\lbrace \tau\in \left\lbrace t_n,\dots,T\right\rbrace \text{ s.t. } \varphi\left(S^1_\tau \right)\geq \bar{p}_n\left(S^1_{\tau},I_{\tau},V_{\tau} \right)   \right\rbrace.
	\end{equation}
	and so we can tackle such a problem again by backward induction. In particular, the hedging strategy is computed by solving the following problem
	\begin{multline}
		\left(\bar{\delta}_{n}^{1},\bar{\delta}_{n}^{2}\right)=	\argmin_{\left(x,y\right)\in \mathbb{R}^2}\rho\left(-\varphi\left(S^1_{\tau_n} \right)+(\bar{\delta}\cdot S)_{n+1}^{{\tau_n}/\Delta t}+\right.\\  	\left.\left[ x\left(S_{t_{n+1}}^{1}-S_{t_{n}}^{1}e^{r\Delta t}\right)+
		y\left(S_{t_{n+1}}^{2}-S_{t_{n}}^{2}e^{r\Delta t}\right)\right]e^{r \left( \tau_n-t_{n+1}\right)}  \right).
	\end{multline}
	Again, such a problem is solved backward in time, by exploiting  Monte Carlo simulations and linear interpolation. {In particular, the calculation of the exercise strategy   requires the comparison between the  payoff and option price, so, in this case, linear interpolation is used for both interpolating the hedging strategy and the option price {at each trading time.}}
	
	\subsection{Transaction costs} Let us consider proportional transaction on traded assets as in Buehler et al. \cite{buehler2019deep}. The inclusion of transaction costs complicates the evaluation of the hedging strategy because the optimal strategy at one instant is closely related to the strategy {employed at the earlier} time-step.
	
	We start by considering the problem of hedging a European derivative, based on the risk measure $\rho$, when proportional transaction costs are due. The following equation defines the Backward Hedging strategy at time $t_n$: 
	\begin{align}
		\left(\bar{\delta}_{n}^{1},\bar{\delta}_{n}^{2}\right)=	\argmin_{\left(x,y\right)\in \mathbb{R}^2}\rho\left(\right.&\left.-\varphi\left(S^1_T \right)+(\bar{\delta}\cdot S)_{n+1}^{N}+C_{n+1}^{N}(\bar{\delta}) \right.\\
		&+\left[ x\left(S_{t_{n+1}}^{1}-S_{t_{n}}^{1}e^{r\Delta t}\right)+
		y\left(S_{t_{n+1}}^{2}-S_{t_{n}}^{2}e^{r\Delta t}\right)\right] e^{\left(T-t_{n+1}\right)r} \notag\\
		&\left.+\varepsilon \left[  S_{t_n}^{1} \left|x- {\delta}_{n-1}^{1}\right| +S_{t_n}^{2} \left|y- {\delta}_{{n-1}}^{2}\right| \right] e^{\left(T-t_n\right)r}  \right)\notag.\label{CT2}
	\end{align}
	Similarly to what we did for the no-transaction-costs case, we replace conditioning with respect to $ \mathcal{F}_{t_n}$ with a discrete set of data, i.e. $S^1_{t_n}=s,I_{t_n}=i, V_{t_n}=v,  {\delta}^1_{{n-1}}=d^1,  {\delta}^2_{{n-1}}=d^2$.
	
	We stress out that, in this case, we cannot eliminate the formula's dependence on $\left( {\delta}_{n-1}^{1}, {\delta}_{n-1}^{2}\right)$, which, proceeding backward, { will be computed at the next time-step as $\left( \bar{\delta}_{n-1}^{1}, \bar{\delta}_{n-1}^{2}\right)$, with the only exception for $ {\delta}_0^1= {\delta}_0^1=0$}. So, we cannot apply the backward approach directly.
	
	To overcome this obstacle, we consider a 5 dimensional grid, which discretizes the conditioning event. Specifically, for each time-step $t_n$, we consider a grid of values $\mathcal{G}_n=\mathcal{G}_n^S\times \mathcal{G}^I_n\times \mathcal{G}^V_n\times \mathcal{G}^{\delta^1}_n\times \mathcal{G}^{\delta^2}_n$ which is defined as follows: $\mathcal{G}_n^S$, $ \mathcal{G}^I_n$ and $\mathcal{G}^V_n$ are defined as in Section \ref{ss1}, while $\mathcal{G}^{\delta^1}_n$ and $ \mathcal{G}^{\delta^2}_n$ are two uniform mesh of $N_{\delta^1}$ and $N_{\delta^2}$ points that represent the values for $ {\delta}_{n-1}^{1}$ and $ {\delta}_{n-1}^{2}$ respectively. Specifically, the maximum and minimum values of $ \mathcal{G}^{\delta^1}_n$ and $ \mathcal{G}^{\delta^2}_n$ are determined as specific quantiles (in our tests we consider $0.1\%$ and $99.9\%$ quantiles) of  $\delta^1_{n-1}$ and  $\delta^2_{n-1}$ {when $\varepsilon=0$}. {Basically}, those quantiles are computed by exploiting Monte Carlo simulations of the Backward Hedging algorithm for the case without transaction costs, starting from time $t=0$.
	The interval for the grids $\mathcal{G}^{\delta^1}_n$ and $ \mathcal{G}^{\delta^2}_n$ are appropriately extended to include the zero value, which is the delta value at the beginning and at the end of the hedging strategy.
	
	The Backward Hedging algorithm for this case is developed following the same valuation principle adopted for the case without transaction costs: it proceeds backward in time and, at each time-step $t_n$, for each point of the 5-dimensional, it determines the optimal values of $\delta^1_n$ and $\delta^2_n$ through a series of Monte Carlo simulations up to maturity (or to the exercise time {with respect to} American options), considering the values associated with the starting grid point as the market state parameters. Again, linear interpolation is used to extend the hedging strategy from the grid nodes to the entire $\mathbb{R}^5$ space. Furthermore, in order to favour the stability of the algorithm in the case of extreme trajectories (that is simulated values are not inside in the grid), the values predicted by the interpolation for $\delta^1_k$ and $\delta^2_k$ for $k=n+1,\dots,N$, are restricted to the discretization interval of $\mathcal{G}^{ {\delta}^1}_k$ and $\mathcal{G}^{ {\delta}^2}_k$ respectively.
	
	\begin{rem}
		\emph{In all cases considered in the preceding sections, for each time-step, the calculation of the {hedging} strategies for each grid point are independent of each other, so this step can be parallelised, thus reducing the total calculation time of the Backward Hedging algorithm.}
	\end{rem}
	
	\section{Numerical results} 
	In this Section we discuss some numerical results related to the proposed algorithm. Specifically, we compute the optimal hedging strategy for a European or American put option, considering either the minimization of the mean square error of hedging or the CVaR at maturity as the objective, including (i.e. $\varepsilon=1\%$) or not transaction costs (i.e. $\varepsilon=0$). Moreover, we consider different settings for $N$, the number of trading dates, and $M$, the number of Monte Carlo simulations used for solving the optimization problems that define the hedging strategy. In particular, in order to obtain reasonable calculation times, we have considered values $M=10^3,10^4,10^5$.
	We first consider the model parameters employed for a test by Buehler et al. \cite{buehler2019deep}, and secondly the same parameters but with positive interest rate and drift. The employed  values are reported in Table \ref{tab:He} together with the put option parameters, while Table \ref{tab:He-1} displays the parameters for the Backward Hedging algorithm.
	
	The following tables show the hedging prices {at time $t_0$} in the various cases considered, obtained by means of the Backward Hedging algorithm (BH), which has been implemented in the MATLAB environment. For the sake of comparison, these prices were also calculated using our implementation of the Deep Hedging algorithm (DH), which has been implemented in Python by us. The parameters for the Deep Hedging algorithm are listed in Table \ref{tab:DH}. We point out that, for the Deep Hedging algorithm, there are two parameters that govern the number of Monte Carlo simulations: $M$ the number of simulations for training the neural network and $M_1$ the number of out-of-sample simulations performed at the end to generate the option price. For the Backward Hedging algorithm, to these two parameters, we add $M_0$, the number of simulations used to determine the grid that discretizes the market states.  For the sake of comparison, we also report the computational times, although these times should be taken with some detachment as they refer to codes developed in different environments. Furthermore, the implementation of the Deep Hedging algorithm was carried out by us and there is probably room for improvement.
	
	The timed comparisons were carried out on a personal computer equipped with an Intel i5-1035G1 processor and 8 GB of RAM, and only one core was used in order to stabilize the calculation times as much as possible. 
	
{	We emphasise that the prices shown in the following tables are determined by considering the profit and loss of the hedging strategy from time zero to maturity, thus representing an opportunity to compare the two hedging strategies on the same objective functions, $\rho\left(PL_0^T \right)$ and $\mathbb{V}\mathrm{ar}\left(PL_0^T \right)$.
 In this regard, DH's strategy should be the optimum point in the set of hedging strategies for the global problem, while BH's strategy can be considered as a solution to the same problem, which includes  the constraint of given by the recursive scheme that defines the backward strategy. Then, through these numerical tests, we can compare the two strategies and see how optimal they are for the global problem.}
	
	\begin{table}[H]
		\begin{centering}
			\begin{tabular}{rlc}
				\toprule  
				Symbol & Meaning & Value\tabularnewline
				\midrule
				$S_{0}$ & Initial spot value & $100$\tabularnewline
				$r$ & Risk free i.r. & $0, 0.1$\tabularnewline
				$\mu$ & Drift & $0, 0.1$\tabularnewline
				$v_{0}$ & Vol at beginning & $0.04$\tabularnewline
				$\alpha$ & Mean reversion speed & $1$\tabularnewline
				$b$ & Long run vol & $0.04$\tabularnewline
				$\sigma$ & Vol of Vol & $2$\tabularnewline
				$\rho_{BW}$ & Correlation & $-0.7$\tabularnewline
				$T$ & Maturity & $1.0$\tabularnewline
				$K$ & Strike & $100$\tabularnewline 
				\bottomrule
			\end{tabular}
			\par\end{centering}
		\caption{\label{tab:He} Parameters of the Heston model and of the put option.}
	\end{table}
	
	\begin{table}[H]
		\begin{centering}
			\begin{tabular}{rlc}
				\toprule  
				Symbol & Meaning & Value\tabularnewline
				\midrule
				$M_{0}$ & Number of MC paths for computing quantiles & $10^{6}$\tabularnewline
				$M$ & Number of MC paths for backward step & $10^3,10^4,10^5$\tabularnewline
				$M_{1}$ & Number of MC paths for out-of-sample step & $10^{6}$\tabularnewline
				$N$ & Number of trading dates & $4,8,16$\tabularnewline
				$N_{T}$ & Number of time-steps for simulation & $16$\tabularnewline
				$N_{S}$ & Number of points to discretyze $S$ & $15$\tabularnewline
				$N_{I}$ &  Number of points  to discretyze the integral of $V$ & $2$\tabularnewline
				$N_{\delta^1}$ &  Number of points  to discretyze $\delta^1_{{n-1}}$ & $3$\tabularnewline
				$N_{\delta^2}$ &  Number of points  to discretyze $\delta^2_{{n-1}}$ & $3$\tabularnewline
				\bottomrule
			\end{tabular}
			\par\end{centering}
		\caption{\label{tab:He-1} Backward Hedging parameters.}
	\end{table}

	\begin{table}[H]
		\begin{centering}
			\begin{tabular}{rlc}
				\toprule  
				Symbol & Meaning & Value\tabularnewline
				\midrule
				$M_{0}$ & Number of MC paths for computing quantiles & $10^{6}$\tabularnewline
				$M$ & Number of MC paths for backward step & $10^3,10^4,10^5$\tabularnewline 
				$N$ & Number of trading dates & $4,8,16$\tabularnewline
				$N_{T}$ & Time-steps for simulation & $64$\tabularnewline
				$N_{N}$ & Number of neurons per layer & $17$\tabularnewline
				$N_{L}$ & Number of hidden layers & $2$\tabularnewline
				$\sigma(x)$ & Activation function & $\max\left(x,0 \right) $\tabularnewline
				\bottomrule
			\end{tabular}
			\par\end{centering}
		\caption{\label{tab:DH} Deep Hedging parameters.}
	\end{table}

	\label{sec:experiments}
	\subsection{Mean squared error} \label{subsec:MSE}
	In this Section we consider the minimization of the mean squered hedging error as target.
	Tables \ref{tab:MSE_EU1} and \ref{tab:MSE_AM1} show the results for the European and the American put option for $r=\mu=0$, while Tables \ref{tab:MSE_EU2} and \ref{tab:MSE_AM2}  report the corresponding results for the  $r=\mu=0.1$ case. We also show the price in Heston's continuous model, obtained through the Hybrid Tree-PDE algorithm by Briani et al. \cite{briani2017hybridapproach,briani2017hybridtree}. This price serves as a benchmark for the case where transaction costs are zero. No benchmark is available for positive transaction costs.
	
	In all the  considered cases, the prices returned by th BH algorithm are very stable  with respect to $M$ and the results are consistent with both the benchmarks (when available) and the results returned by the DH algorithm (for the European option).  When $r=\mu=0$, the American price is very similar to the European price, but when $r=\mu=0.10$, the difference is clear. In the latter case, we observe that the price increases as the number of instants of hedging increases because the number of possibilities for exercising the option increases.

	To conclude, the proposed algorithm proves to be very efficient and stable.
	\begin{table}[H]
		\begin{centering}
			\scalebox{0.8}{
				\setlength{\tabcolsep}{4pt} 
				\renewcommand{\arraystretch}{1} 
				\begin{tabular}{ccrccccccc}
					\toprule 
					\multicolumn{3}{c}{ } & \multicolumn{3}{c}{$\varepsilon=0$} &  & \multicolumn{3}{c}{$\varepsilon=1\%$}\tabularnewline
					\cmidrule{4-10} \cmidrule{5-10} \cmidrule{6-10} \cmidrule{7-10} \cmidrule{8-10} \cmidrule{9-10} \cmidrule{10-10} 
					\multicolumn{3}{c}{Benchmark} & \multicolumn{3}{c}{$3.7165$} &  &  & $-$ & \tabularnewline
					& $M$ & $N$ & $4$ & $8$ & $16$ &  & $4$ & $8$ & $16$\tabularnewline
					\midrule
					& $10^{3}$ &  & $\underset{\left(7\right)}{3.7361}$ & $\underset{\left(10\right)}{3.7518}$ & $\underset{\left(19\right)}{3.8075}$ &  & $\underset{\left(97\right)}{5.2907}$ & $\underset{\left(377\right)}{5.6050}$ & $\underset{\left(1501\right)}{7.0452}$\tabularnewline
					BH & $10^{4}$ &  & $\underset{\left(8\right)}{3.7024}$ & $\underset{\left(15\right)}{3.7173}$ & $\underset{\left(32\right)}{3.6962}$ &  & $\underset{\left(359\right)}{5.1429}$ & $\underset{\left(2059\right)}{5.3127}$ & $\underset{\left(8841\right)}{5.4350}$\tabularnewline
					& $10^{5}$ &  & $\underset{\left(16\right)}{3.7081}$ & $\underset{\left(44\right)}{3.7031}$ & $\underset{\left(130\right)}{3.7061}$ &  & $\underset{\left(2548\right)}{5.0779}$ & $\underset{\left(16919\right)}{5.1680}$ & $\underset{\left(75182\right)}{5.2412}$\tabularnewline
					\midrule
					& $10^{3}$ &  & $\underset{\left(200\right)}{3.3484}$ & $\underset{\left(622\right)}{2.6859}$ & $\underset{\left(721\right)}{2.6354}$ &  & $\underset{\left(172\right)}{5.3223}$ & $\underset{\left(624\right)}{6.3477}$ & $\underset{\left(1188\right)}{8.0656}$\tabularnewline
					DH & $10^{4}$ &  & $\underset{\left(991\right)}{3.6130}$ & $\underset{\left(2153\right)}{3.5731}$ & $\underset{\left(4152\right)}{3.4796}$ &  & $\underset{\left(2752\right)}{5.4080}$ & $\underset{\left(3007\right)}{6.9781}$ & $\underset{\left(4918\right)}{10.5884}$\tabularnewline
					& $10^{5}$ &  & $\underset{\left(8741\right)}{3.7019}$ & $\underset{\left(19587\right)}{3.7241}$ & $\underset{\left(51194\right)}{3.7426}$ &  & $\underset{\left(18890\right)}{5.3400}$ & $\underset{\left(50125\right)}{6.6993}$ & $\underset{\left(92425\right)}{9.3661}$\tabularnewline
					\bottomrule
			\end{tabular}}
			\par\end{centering}
		\caption{\label{tab:MSE_EU1}Prices for a European put option with
			$r=\mu=0$, while minimizing loss based on mean squared error. }
	\end{table}
	
	\begin{table}[H]
		\begin{centering}
			\scalebox{0.8}{
				\setlength{\tabcolsep}{4pt} 
				\renewcommand{\arraystretch}{1} 
				\begin{tabular}{ccrccccccc}
					\toprule 
					\multicolumn{3}{c}{ } & \multicolumn{3}{c}{$\varepsilon=0$} &  & \multicolumn{3}{c}{$\varepsilon=1\%$}\tabularnewline
					\cmidrule{4-10} \cmidrule{5-10} \cmidrule{6-10} \cmidrule{7-10} \cmidrule{8-10} \cmidrule{9-10} \cmidrule{10-10} 
					\multicolumn{3}{c}{Benchmark} &  & $3.7165$ &  &  &  & $-$ & \tabularnewline
					& $M$ & $N$ & $4$ & $8$ & $16$ &  & $4$ & $8$ & $16$\tabularnewline
					\midrule
					& $10^{3}$ &  & $\underset{\left(8\right)}{3.7360}$ & $\underset{\left(16\right)}{3.7513}$ & $\underset{\left(49\right)}{3.8066}$ &  & $\underset{\left(86\right)}{5.2907}$ & $\underset{\left(382\right)}{5.6603}$ & $\underset{\left(1618\right)}{6.9647}$\tabularnewline
					BH & $10^{4}$ &  & $\underset{\left(12\right)}{3.7022}$ & $\underset{\left(35\right)}{3.7172}$ & $\underset{\left(111\right)}{3.6958}$ &  & $\underset{\left(407\right)}{5.0861}$ & $\underset{\left(2377\right)}{5.3101}$ & $\underset{\left(10195\right)}{5.5282}$\tabularnewline
					& $10^{5}$ &  & $\underset{\left(42\right)}{3.7081}$ & $\underset{\left(197\right)}{3.7038}$ & $\underset{\left(872\right)}{3.7057}$ &  & $\underset{\left(3336\right)}{5.1669}$ & $\underset{\left(20906\right)}{5.1404}$ & $\underset{\left(88598\right)}{5.3706}$\tabularnewline
					\bottomrule
			\end{tabular}}
			\par\end{centering}
		\caption{\label{tab:MSE_AM1}Prices for an American put option with
			$r=\mu=0$, while minimizing loss based on mean squared error.}
	\end{table}
	
	\begin{table}[H]
		\begin{centering}
			\scalebox{0.8}{
				\setlength{\tabcolsep}{4pt} 
				\renewcommand{\arraystretch}{1} 
				\begin{tabular}{ccrccccccc}
					\toprule 
					\multicolumn{3}{c}{ } & \multicolumn{3}{c}{$\varepsilon=0$} &  & \multicolumn{3}{c}{$\varepsilon=1\%$}\tabularnewline
					\cmidrule{4-10} \cmidrule{5-10} \cmidrule{6-10} \cmidrule{7-10} \cmidrule{8-10} \cmidrule{9-10} \cmidrule{10-10} 
					\multicolumn{3}{c}{Benchmark} &  & $2.1334$ &  &  &  &$-$& \tabularnewline
					& $M$ & $N$ & $4$ & $8$ & $16$ &  & $4$ & $8$ & $16$\tabularnewline
					\midrule
					& $10^{3}$ &  & $\underset{\left(6\right)}{2.1537}$ & $\underset{\left(10\right)}{2.1470}$ & $\underset{\left(18\right)}{2.2142}$ &  & $\underset{\left(96\right)}{2.9850}$ & $\underset{\left(391\right)}{3.1667}$ & $\underset{\left(1499\right)}{3.9847}$\tabularnewline
					BH & $10^{4}$ &  & $\underset{\left(8\right)}{2.1392}$ & $\underset{\left(14\right)}{2.1547}$ & $\underset{\left(31\right)}{2.1412}$ &  & $\underset{\left(364\right)}{3.0313}$ & $\underset{\left(2072\right)}{3.0702}$ & $\underset{\left(8895\right)}{3.0806}$\tabularnewline
					& $10^{5}$ &  & $\underset{\left(16\right)}{2.1462}$ & $\underset{\left(43\right)}{2.1401}$ & $\underset{\left(147\right)}{2.1410}$ &  & $\underset{\left(2674\right)}{2.9695}$ & $\underset{\left(17137\right)}{2.8845}$ & $\underset{\left(74737\right)}{2.9150}$\tabularnewline
					\midrule 
					\multirow{3}{*}{DH} & $10^{3}$ &  & $\underset{\left(208\right)}{1.3124}$ & $\underset{\left(396\right)}{1.3044}$ & $\underset{\left(677\right)}{1.1886}$ &  & $\underset{\left(440\right)}{2.3709}$ & $\underset{\left(506\right)}{3.9069}$ & $\underset{\left(800\right)}{5.2301}$\tabularnewline
					& $10^{4}$ &  & $\underset{\left(944\right)}{1.9725}$ & $\underset{\left(1981\right)}{1.9670}$ & $\underset{\left(4399\right)}{1.7516}$ &  & $\underset{\left(900\right)}{3.4638}$ & $\underset{\left(2427\right)}{4.7569}$ & $\underset{\left(4678\right)}{4.6196}$\tabularnewline
					& $10^{5}$ &  & $\underset{\left(11319\right)}{2.1270}$ & $\underset{\left(20367\right)}{2.0983}$ & $\underset{\left(41625\right)}{2.1339}$ &  & $\underset{\left(27554\right)}{3.2011}$ & $\underset{\left(64813\right)}{3.9313}$ & $\underset{\left(85591\right)}{4.1068}$\tabularnewline
					\bottomrule
			\end{tabular}}
			\par\end{centering}
		\caption{\label{tab:MSE_EU2}Prices for a European put option with
			$r=\mu=0.1$, while minimizing loss based on mean squared error.}
	\end{table}
	\begin{table}[H]
		\begin{centering}
			\scalebox{0.8}{
				\setlength{\tabcolsep}{4pt} 
				\renewcommand{\arraystretch}{1} 
				\begin{tabular}{ccrccccccc}
					\toprule 
					\multicolumn{3}{c}{ } & \multicolumn{3}{c}{$\varepsilon=0$} &  & \multicolumn{3}{c}{$\varepsilon=1\%$}\tabularnewline
					\cmidrule{4-10} \cmidrule{5-10} \cmidrule{6-10} \cmidrule{7-10} \cmidrule{8-10} \cmidrule{9-10} \cmidrule{10-10} 
					\multicolumn{3}{c}{Benchmark} &  & $2.8355$ &  &  &  & $-$ & \tabularnewline
					& $M$ & $N$ & $4$ & $8$ & $16$ &  & $4$ & $8$ & $16$\tabularnewline
					\midrule
					& $10^{3}$ &  & $\underset{\left(6\right)}{2.6580}$ & $\underset{\left(10\right)}{2.7427}$ & $\underset{\left(20\right)}{2.8339}$ &  & $\underset{\left(87\right)}{3.5394}$ & $\underset{\left(387\right)}{3.8889}$ & $\underset{\left(1423\right)}{4.7330}$\tabularnewline
					BH & $10^{4}$ &  & $\underset{\left(8\right)}{2.6238}$ & $\underset{\left(16\right)}{2.7503}$ & $\underset{\left(39\right)}{2.7745}$ &  & $\underset{\left(444\right)}{3.4659}$ & $\underset{\left(2945\right)}{3.8864}$ & $\underset{\left(10535\right)}{3.7912}$\tabularnewline
					& $10^{5}$ &  & $\underset{\left(20\right)}{2.6264}$ & $\underset{\left(64\right)}{2.7376}$ & $\underset{\left(203\right)}{2.7730}$ &  & $\underset{\left(4140\right)}{3.5296}$ & $\underset{\left(27944\right)}{3.5533}$ & $\underset{\left(93940\right)}{3.6406}$\tabularnewline
					\bottomrule
			\end{tabular}}
			\par\end{centering}
		\caption{\label{tab:MSE_AM2}Prices for an American put option with
			$r=\mu=0.1$, while minimizing loss based on mean squared error.}
	\end{table}
	\FloatBarrier
	\subsection{CVaR error}  
	
	In this Section we consider the minimization of the {$95\%$}-CVaR of the hedging strategy at maturity as target. {Before discussing the results, we point out that, in order to facilitate comparison between the proposed algorithms, in this case we have also included, for the DH algorithm only, the case $M=10^6$, which is necessary to obtain competitive results with BH.}\\
	Tables \ref{tab:CVaR_EU1} and \ref{tab:CVaR_AM1} show the results for the European and the American put option for $r=\mu=0$, while Tables \ref{tab:CVaR_EU2} and \ref{tab:CVaR_AM2}  report the corresponding results for the  $r=\mu=0.1$ case.   No benchmark is available in this case.
	In all cases considered, the prices returned by th BH algorithm are very stable  with respect to $M$. As far as European options are considered, the  price decreases as $N$, the number of possible portfolio rebalances, increases, since the higher this number, the smaller the error committed at the end of the hedging period and the lower the relative CVaR. In fact, when one aims to minimise a risk measure of replication error, unlike mean squares error minimisation, the lower the price obtained, the better the strategy.

	As observed in the previous numerical experiments, discussed in Subsection \ref{subsec:MSE}, in the presence of transaction costs, the option price increases, as expected. As noted in the previous section, when $r=\mu=0$, the American price is very similar to the European price, but when $r=\mu=0.10$, the difference is clear. In the latter case, we observe that the price decreases: while the increase in trading dates favours the optimal exercise of the option that raises the price, it also favours the development of an effective hedging strategy that lowers the price. Between these two contrasting effects, the second seems to prevail. 
	{As shown in Tables \ref{tab:CVaR_EU1} and \ref{tab:CVaR_EU2}, the comparison with DH shows that BH's proposed strategy is also very efficient in dealing with the global problem:  {for all the parameter configurations considered,} the price of the BH strategy is very similar to the price of DH's strategy, {as their difference is small}, and in some cases even lower. Moreover, the computational time tends to reward BH.} {We also observe that the run times of BH are more regular than those of DH, as the  stop criterion employed for training the neural network seems to be more sensitive to specific input data. In particular, DH times seem to increase up to $M=10^5$, whereas for $M=10^6$ they are of the same order of magnitude as for $M=10^5$.}	
	In conclusion, BH again proves to be very efficient and stable.
	
\begin{figure}
		\begin{center}
			\includegraphics[width=0.9\textwidth]{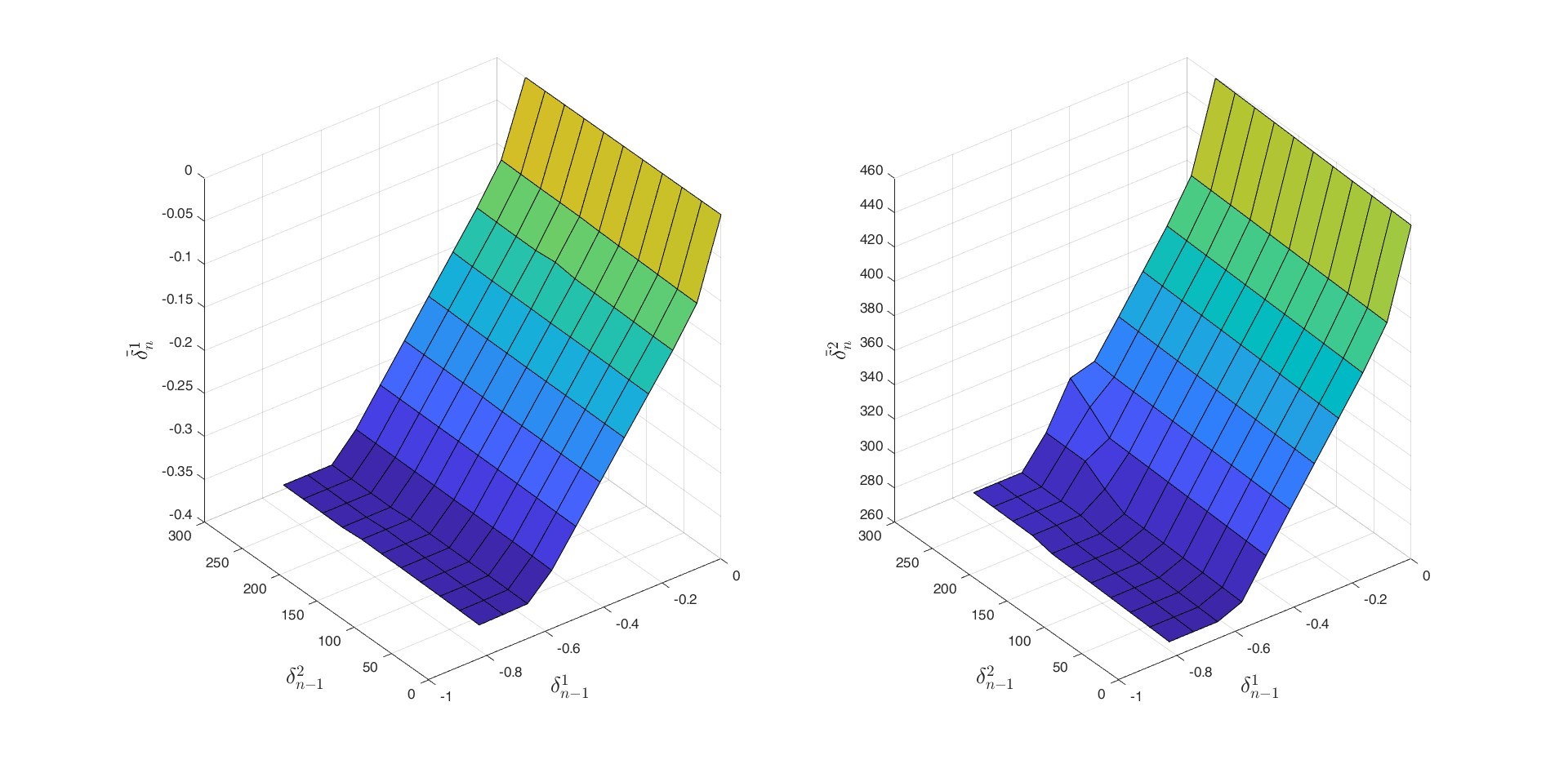} 
		\end{center}
		\caption{\label{F1}Optimal values of $\bar{\delta}^1_3$ (right) and  $\bar{\delta}^2_3$ (left) as a function of $\left(\delta^1_2,\delta^2_2 \right)$.}
	\end{figure}

	To complete our analysis, we present Figure \ref{F1} which shows a plot of the optimal values of $\bar{\delta}^1_3$  and  $\bar{\delta}^2_3$, the hedging strategy at time $t_3=0.75$, computed through the Backward Hedging algorithm, as a function of the hedging strategy at time $t_2=0.5$, $\left( \delta^1_2,\delta^2_2\right) $. In particular, we consider an American put option with  $N=4$ hedging steps, $S_{0.75}=100$, $I_{0.75}=0.04\cdot0.75$ and $V_{0.75}=0.04$, with $M=10^4$ Monte Carlo simulations. Hedging is based on the CVaR minimization and it employs $\varepsilon=1\%$ proportional transaction cost. To improve the graphic detail, the values $N_{\delta_1}$ and $N_{\delta_2}$ have been set equal to $11$.
	
\begin{figure}
		\begin{center} 
			\includegraphics[width=0.9\textwidth]{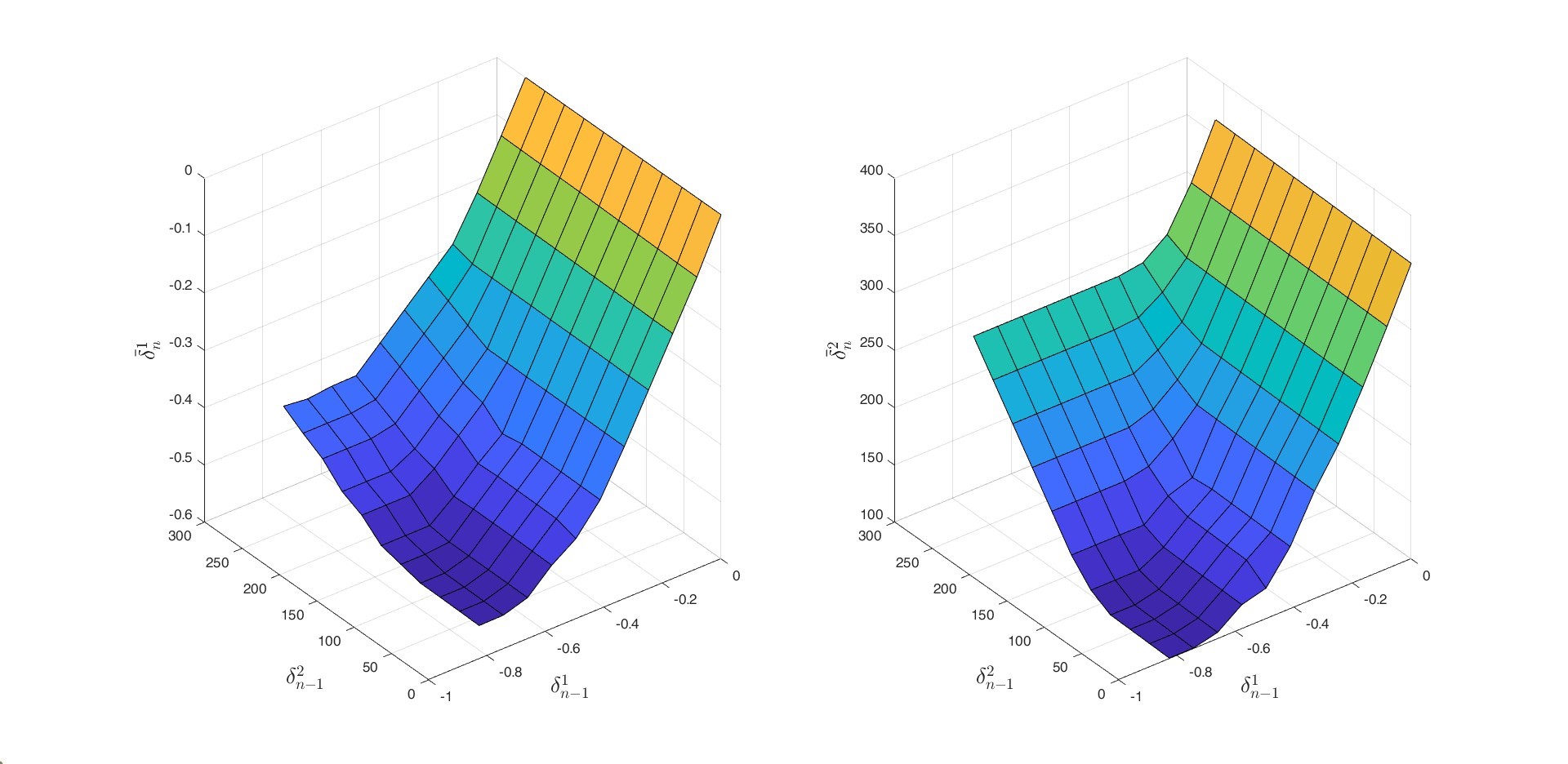}
		\end{center}
		\caption{\label{F2}Optimal values of $\bar{\delta}^1_3$ (right) and  $\bar{\delta}^2_3$ (left) as a function of $\left(\delta^1_2,\delta^2_2 \right)$. Proportional cost for trading $S^1$ is $1\%$, while the proportional cost for trading $S^2$ has been raised to $10\%$.}
	\end{figure}

	The graphs suggest that both $\bar{\delta}^1_3$  and  $\bar{\delta}^2_3$ {increase} as $\delta^1_2$  and  $\delta^2_2$ increase, with a greater dependence on $\delta^1_2$. This fact probably results from the low absolute cost in trading $S^2$ due to the small price of $S^2$ with respect of $S^1$. To confirm this, {in Figure} \ref{F2} we can observe a graph similar to that of {Figure} \ref{F1}, in which the relative cost of trading $S^2$ has been raised to $10\%$, while that for $S^1$ has remained $1\%$. In this case, two increasing graphs are still observed, but the dependence on $\delta^1_2$ is more relevant, in particular for $\bar{\delta}^2_3$.
	
	\begin{table}[H]
		\begin{centering}
			\scalebox{0.8}{
				\setlength{\tabcolsep}{4pt} 
				\renewcommand{\arraystretch}{1} 
				\begin{tabular}{ccrccccccc}
					\toprule 
					\multicolumn{3}{c}{ } & \multicolumn{3}{c}{$\varepsilon=0$} &  & \multicolumn{3}{c}{$\varepsilon=1\%$}\tabularnewline
					\cmidrule{4-10} \cmidrule{5-10} \cmidrule{6-10} \cmidrule{7-10} \cmidrule{8-10} \cmidrule{9-10} \cmidrule{10-10} 
					&  &  &  &  &  &  &  &  & \tabularnewline
					& $M$ & $N$ & $4$ & $8$ & $16$ &  & $4$ & $8$ & $16$\tabularnewline
					\midrule
					& $10^{3}$ &  & $\underset{\left(10\right)}{10.4559}$ & $\underset{\left(14\right)}{8.7008}$ & $\underset{\left(24\right)}{7.6950}$ &  & $\underset{\left(133\right)}{11.5441}$ & $\underset{\left(502\right)}{10.3812}$ & $\underset{\left(1978\right)}{9.3263}$\tabularnewline
					BH & $10^{4}$ &  & $\underset{\left(35\right)}{10.1111}$ & $\underset{\left(74\right)}{8.6460}$ & $\underset{\left(143\right)}{7.4112}$ &  & $\underset{\left(467\right)}{11.5375}$ & $\underset{\left(2510\right)}{10.3645}$ & $\underset{\left(10388\right)}{9.3284}$\tabularnewline
					& $10^{5}$ &  & $\underset{\left(351\right)}{10.0945}$ & $\underset{\left(762\right)}{8.6725}$ & $\underset{\left(1566\right)}{7.4855}$ &  & $\underset{\left(3440\right)}{11.5099}$ & $\underset{\left(19424\right)}{10.3386}$ & $\underset{\left(80715\right)}{9.3564}$\tabularnewline
					\midrule
					& $10^{3}$ &  & $\underset{\left(704\right)}{20.7569}$ & $\underset{\left(794\right)}{22.8564}$ & $\underset{\left(1908\right)}{45.1245}$ &  & $\underset{\left(748\right)}{25.8685}$ & $\underset{\left(305\right)}{25.4389}$ & $\underset{\left(500\right)}{24.9047}$\tabularnewline
					DH & $10^{4}$ &  & $\underset{\left(2001\right)}{18.4303}$ & $\underset{\left(1539\right)}{20.9419}$ & $\underset{\left(5676\right)}{24.5623}$ &  & $\underset{\left(1748\right)}{20.3021}$ & $\underset{\left(784\right)}{27.6064}$ & $\underset{\left(1455\right)}{25.3484}$\tabularnewline
					& $10^{5}$ &  & $\underset{\left(32783\right)}{12.1416}$ & $\underset{\left(52974\right)}{10.7884}$ & $\underset{\left(45811\right)}{12.6447}$ &  & $\underset{\left(4667\right)}{11.7335}$ & $\underset{\left(9600\right)}{12.0040}$ & $\underset{\left(11239\right)}{13.7020}$\tabularnewline
					& $10^{6}$ &  & $\underset{\left(21818\right)}{9.9972}$ & $\underset{\left(42418\right)}{9.2974}$ & $\underset{\left(84488 \right)}{8.3096 }$ &  & $\underset{\left(28314\right)}{11.5243 }$ & $\underset{\left(45034 \right)}{9.9993 }$ & $\underset{\left(100586\right)}{ 9.0129}$\tabularnewline
					\bottomrule
			\end{tabular}}
			\par\end{centering}
		\caption{\label{tab:CVaR_EU1}Hedging cost for a European put option with
			$r=\mu=0$, while minimizing loss based on CVaR.}
	\end{table}
	\begin{table}[H]
		\begin{centering}
			\scalebox{0.8}{
				\setlength{\tabcolsep}{4pt} 
				\renewcommand{\arraystretch}{1} 
				\begin{tabular}{ccrccccccc}
					\toprule 
					\multicolumn{3}{c}{ } & \multicolumn{3}{c}{$\varepsilon=0$} &  & \multicolumn{3}{c}{$\varepsilon=1\%$}\tabularnewline
					\cmidrule{4-10} \cmidrule{5-10} \cmidrule{6-10} \cmidrule{7-10} \cmidrule{8-10} \cmidrule{9-10} \cmidrule{10-10} 
					&  &  &  &  &  &  &  &  & \tabularnewline
					& $M$ & $N$ & $4$ & $8$ & $16$ &  & $4$ & $8$ & $16$\tabularnewline
					\midrule
					& $10^{3}$ &  & $\underset{\left(10\right)}{10.4560}$ & $\underset{\left(14\right)}{8.7006}$ & $\underset{\left(27\right)}{7.6962}$ &  & $\underset{\left(119\right)}{11.5435}$ & $\underset{\left(499\right)}{10.3831}$ & $\underset{\left(2087\right)}{9.3273}$\tabularnewline
					BH & $10^{4}$ &  & $\underset{\left(35\right)}{10.1111}$ & $\underset{\left(73\right)}{8.6462}$ & $\underset{\left(148\right)}{7.4082}$ &  & $\underset{\left(476\right)}{11.5382}$ & $\underset{\left(2733\right)}{10.3588}$ & $\underset{\left(11618\right)}{9.3308}$\tabularnewline
					& $10^{5}$ &  & $\underset{\left(355\right)}{10.0944}$ & $\underset{\left(759\right)}{8.6724}$ & $\underset{\left(1625\right)}{7.4855}$ &  & $\underset{\left(3721\right)}{11.5117}$ & $\underset{\left(21980\right)}{10.3395}$ & $\underset{\left(91381\right)}{9.3630}$\tabularnewline
					\bottomrule
			\end{tabular}}
			\par\end{centering}
		\caption{\label{tab:CVaR_AM1}Hedging cost for an American put option with
			$r=\mu=0$, while minimizing loss based on CVaR.}
	\end{table}
	\begin{table}[H]
		\begin{centering}
			\scalebox{0.8}{
				\setlength{\tabcolsep}{4pt} 
				\renewcommand{\arraystretch}{1} 
				\begin{tabular}{ccrccccccc}
					\toprule 
					\multicolumn{3}{c}{ } & \multicolumn{3}{c}{$\varepsilon=0$} &  & \multicolumn{3}{c}{$\varepsilon=1\%$}\tabularnewline
					\cmidrule{4-10} \cmidrule{5-10} \cmidrule{6-10} \cmidrule{7-10} \cmidrule{8-10} \cmidrule{9-10} \cmidrule{10-10} 
					&  &  &  &  &  &  &  &  & \tabularnewline
					& $M$ & $N$ & $4$ & $8$ & $16$ &  & $4$ & $8$ & $16$\tabularnewline
					\midrule
					& $10^{3}$ &  & $\underset{\left(13\right)}{6.5798}$ & $\underset{\left(13\right)}{5.3969}$ & $\underset{\left(23\right)}{4.4896}$ &  & $\underset{\left(139\right)}{7.4402}$ & $\underset{\left(497\right)}{6.1939}$ & $\underset{\left(1957\right)}{5.3032}$\tabularnewline
					BH & $10^{4}$ &  & $\underset{\left(35\right)}{6.5619}$ & $\underset{\left(70\right)}{5.2388}$ & $\underset{\left(140\right)}{4.3412}$ &  & $\underset{\left(448\right)}{7.3739}$ & $\underset{\left(2443\right)}{6.0801}$ & $\underset{\left(10104\right)}{5.3044}$\tabularnewline
					& $10^{5}$ &  & $\underset{\left(359\right)}{6.5139}$ & $\underset{\left(803\right)}{5.2283}$ & $\underset{\left(1545\right)}{4.3347}$ &  & $\underset{\left(3354\right)}{7.3460}$ & $\underset{\left(20161\right)}{6.0491}$ & $\underset{\left(79368\right)}{5.3076}$\tabularnewline
					\midrule
					& $10^{3}$ &  & $\underset{\left(388\right)}{18.1152}$ & $\underset{\left(522\right)}{23.8998}$ & $\underset{\left(1530\right)}{24.1038}$ &  & $\underset{\left(424\right)}{24.5469}$ & $\underset{\left(802\right)}{27.4145}$ & $\underset{\left(1689\right)}{29.8575}$\tabularnewline
					DH & $10^{4}$ &  & $\underset{\left(1620\right)}{17.5067}$ & $\underset{\left(1818\right)}{19.2940}$ & $\underset{\left(3710\right)}{18.3218}$ &  & $\underset{\left(2031\right)}{18.2993}$ & $\underset{\left(1892\right)}{18.8594}$ & $\underset{\left(5720\right)}{21.2792}$\tabularnewline
					& $10^{5}$ &  & $\underset{\left(26602\right)}{10.2705}$ & $\underset{\left(11261\right)}{12.2695}$ & $\underset{\left(67796\right)}{8.1678}$ &  & $\underset{\left(42343\right)}{10.3475}$ & $\underset{\left(53501\right)}{8.1222}$ & $\underset{\left(105889\right)}{7.9165}$\tabularnewline
					& $10^{6}$ &  & $\underset{\left(22974\right)}{6.6167 }$ & $\underset{\left(31135\right)}{5.8382}$ & $\underset{\left(42442\right)}{5.3762}$ &  & $\underset{\left(19964\right)}{6.9460}$ & $\underset{\left(62982\right)}{5.9089}$ & $\underset{\left(110851\right)}{5.2575}$\tabularnewline
					\bottomrule
			\end{tabular}}
			\par\end{centering}
		\caption{\label{tab:CVaR_EU2}Hedging cost for a European put option with
			$r=\mu=0.1$, while minimizing loss based on CVaR.}
	\end{table}
	\begin{table}[H]
		\begin{centering}
			\scalebox{0.8}{
				\setlength{\tabcolsep}{4pt} 
				\renewcommand{\arraystretch}{1} 
				\begin{tabular}{ccrccccccc}
					\toprule 
					\multicolumn{3}{c}{ } & \multicolumn{3}{c}{$\varepsilon=0$} &  & \multicolumn{3}{c}{$\varepsilon=1\%$}\tabularnewline
					\cmidrule{4-10} \cmidrule{5-10} \cmidrule{6-10} \cmidrule{7-10} \cmidrule{8-10} \cmidrule{9-10} \cmidrule{10-10} 
					&  &  &  &  &  &  &  &  & \tabularnewline
					& $M$ & $N$ & $4$ & $8$ & $16$ &  & $4$ & $8$ & $16$\tabularnewline
					\midrule
					& $10^{3}$ &  & $\underset{\left(9\right)}{8.2635}$ & $\underset{\left(14\right)}{7.0993}$ & $\underset{\left(30\right)}{6.0820}$ &  & $\underset{\left(117\right)}{ 8.7475}$ & $\underset{\left(438\right)}{7.7933}$ & $\underset{\left(1729\right)}{7.3570}$\tabularnewline
					BH & $10^{4}$ &  & $\underset{\left(34\right)}{8.3280}$ & $\underset{\left(71\right)}{6.9922}$ & $\underset{\left(143\right)}{6.0175}$ &  & $\underset{\left(454\right)}{8.6478}$ & $\underset{\left(2332\right)}{7.9860}$ & $\underset{\left(9742\right)}{7.4585}$\tabularnewline
					& $10^{5}$ &  & $\underset{\left(341\right)}{8.2982}$ & $\underset{\left(731\right)}{6.9864}$ & $\underset{\left(1530\right)}{5.9918}$ &  & $\underset{\left(3539\right)}{8.8270}$ & $\underset{\left(19797\right)}{8.0841}$ & $\underset{\left(77681\right)}{7.5112}$\tabularnewline
					\bottomrule
			\end{tabular}}
			\par\end{centering}
		\caption{\label{tab:CVaR_AM2}Hedging cost for an American put option with
			$r=\mu=0.1$, while minimizing loss based on CVaR.}
	\end{table}
	
	\FloatBarrier
	
	\FloatBarrier
	\section{Conclusions}
	\label{sec:conclusions} 
	In this paper, we have discussed a method for identifying the most efficient hedging strategies for derivative products, including both European and American options, while taking into account transaction costs. The hedging strategy, defined as the quantities of assets held in the hedging portfolio, is determined to minimize a risk measure or the mean squared error of the hedging error at the time the option is exercised.  We calculate the optimal strategy for each point on a grid by using Monte Carlo simulations and then we extend the strategy to other points via interpolation. {Although the proposed strategy is not optimal at time zero for the global optimization problem, it offers, at each time-step, a strategy that is optimal given the strategy that will be followed later, so it is easier to follow in practice, pleasing both practitioners and supervisors.}
	The proposed method proves to be effective and reliable, offering clear interpretation of the results. The algorithm has the potential to be valuable for practitioners managing option positions on a single underlying asset.

	\bibliographystyle{abbrv}
	\bibliography{references}
	
	 \newpage
	\appendix

	\section{Pseudocode} \label{sec:pc}
	This Appendix contains pseudocodes of the Backward Hedging algorithm for the valuation of European and American options.
	\begin{algorithm}
		\caption{Backward Hedging for European options}
		\label{alg:BH}
		{\small 	\begin{algorithmic}   
				\STATE{Set up volatility tree  $\mathcal{T}$ by following \cite{goudenege2021gaussian}}
				\IF{$\varepsilon>0$}
				\STATE{run this algorithm with $\varepsilon=0$ to obtain the parameters for the 5D grids $\mathcal{G}_n$ for $n=1,\dots,N-1$}
				\ELSE
				\STATE{simulate $M_0$ random paths  $\left(  {S}^1_{n,m},  {I}_{n,m}\right) $ for $k=1,\dots,N$}
				\FOR{$n=N-1,\dots,0$}
				\STATE{compute quantiles from $\left( \tilde {S}^1_{n,m}, \tilde{I}_{n,m}\right) $ and set the 3D grid $\mathcal{G}_n$ from quantiles and $\mathcal{T}$}
				\ENDFOR
				\ENDIF
				\FOR{$n=N-1,\dots,0$} 
				\FOR{all points $G$ of $\mathcal{G}_n$ (if $\varepsilon=0$ then $G=(s,i,v)$ else $G=(s,i,v,d^1,d^2)$ )}
				\STATE{simulate $M$ random paths $\left( S^1_{k,m}, S^2_{k,m},I_{k,m},V_{k,m}\right) $ for $k=n+1,\dots,N$, according to $S^1_{t_n}=s,I_{t_n}=i,V_{t_n}=v$}
				
				\WHILE{ the Simplex Search method has not converged}
				\STATE{change $\left(\delta_{n,m}^{1},\delta_{n,m}^{2}\right)$ according to the Simplex Search method}
				\FOR{$m=1,\dots,M$ (for each simulated path)}
				\FOR{$k=n+1,\dots,N$}
				\STATE{ compute the hedging strategy $\left(\bar{\delta}_{k,m}^{1},\bar{\delta}_{k,m}^{2}\right)$} by linear interpolation over $\mathcal{G}_k$ (given $\left(\delta_{k-1,m}^{1},\delta_{k-1,m}^{2}\right)$ if $\varepsilon>0$)
				\ENDFOR
				\STATE{compute the hedging error $- {\varphi\left(S^1_{T,m} \right)}+(\bar{\delta}\cdot S_m)_{n}^{N}+C_{n}^{N}(\bar{\delta}) $} 
				\ENDFOR
				\STATE{estimate $\rho\left(-\varphi\left(S^1_T \right)+\left(\bar{\delta}\cdot S\right)_{n}^{N} +C_{n}^{N}(\bar{\delta}) \right)$ from simulations}
				\ENDWHILE
				\ENDFOR
				\IF{$n>0$}
				\STATE{compute and store coefficients for linear interpolation of $\bar{\delta}^1_n$ and  $\bar{\delta}^2_n$ over $\mathcal{G}_n$}
				\ENDIF
				\ENDFOR
				
				\STATE{Simulate $M_1$ random paths $\left(  {S}^1_{n,m},   {S}^2_{n,m},  {I}_{n,m},  {V}_{n,m}\right) $ for $n=1,\dots,N$} 
				\FOR{$m=1,\dots,M$ (for each simulated path)}
				\STATE{ compute the hedging strategy $\left(\bar{\delta}_{n,m}^{1},\bar{\delta}_{n,m}^{2}\right)$ for $n=0,\dots,N$,  the payoff $ {\varphi\left(S^1_T \right)}_m$, and the hedging error $- {\varphi\left(S^1_T \right)}_m+(\bar{\delta}\cdot  {S}_m)_{n}^{N} +C_{n}^{N}(\bar{\delta})$ } 
				\ENDFOR
				\IF{$\varepsilon>0$}
				\RETURN{estimate $\bar{p}_0=\rho\left(- {\varphi\left(S^1_T \right)}+\left(\bar{\delta}\cdot  {S}\right)_{0}^{N}+C_{0}^{N}(\bar{\delta}) \right)e^{-rT}$}
				\ELSE
				\RETURN{estimate $\bar{p}_0=\rho\left(- {\varphi\left(S^1_T \right)}+\left(\bar{\delta}\cdot  {S}\right)_{0}^{N}\right)e^{-rT}$ and, for $n=1,\dots,N-1$, the quantiles of $S^1_{t_n},I_{t_n},\bar{\delta}^1_n$ and $\bar{\delta}^2_n$} based on the out of sample simulations
				\ENDIF
		\end{algorithmic}}
	\end{algorithm} 
	
	\begin{algorithm}
		\caption{Backward Hedging for American options}
		\label{alg:BHA}
		{\small 	\begin{algorithmic}   
				\STATE{Set up volatility tree  $\mathcal{T}$ by following \cite{goudenege2021gaussian}}
				\IF{$\varepsilon>0$}
				\STATE{run this algorithm with $\varepsilon=0$ to obtain the parameters for the 5D grids $\mathcal{G}_n$ for $n=1,\dots,N-1$}
				\ELSE
				\STATE{simulate $M_0$ random paths  $\left(  {S}^1_{n,m},  {I}_{n,m}\right) $ for $k=1,\dots,N$}
				\FOR{$n=N-1,\dots,0$}
				\STATE{compute quantiles from $\left(  {S}^1_{n,m},  {I}_{n,m}\right) $ and set the 3D grid $\mathcal{G}_n$ from quantiles and $\mathcal{T}$}
				\ENDFOR
				\ENDIF
				\FOR{$n=N-1,\dots,0$} 
				\FOR{all points $G$ of $\mathcal{G}_n$ (if $\varepsilon=0$ then $G=(s,i,v)$ else $G=(s,i,v,d^1,d^2)$ )}
				\STATE{simulate $M$ random paths $\left( S^1_{k,m}, S^2_{k,m},I_{k,m},V_{k,m}\right) $ for $k=n+1,\dots,N$, according to $S^1_{t_n}=s,I_{t_n}=i,V_{t_n}=v$}
				
				\WHILE{ the Simplex Search method has not converged}
				\STATE{change $\left(\delta_{n,m}^{1},\delta_{n,m}^{2}\right)$ according to the Simplex Search method}
				\FOR{$m=1,\dots,M$ (for each simulated path)}
				\STATE{compute worst hedging stopping time $\tau_{n,m}$ by comparing future payoff $\varphi\left(S^1_{k,m} \right) $ and hedging cost $\bar{p}_k$ (estimated by interpolation)}
				\FOR{$k=n+1,\dots,\tau_{n,m}/\Delta t$}
				\STATE{ compute the hedging strategy $\left(\bar{\delta}_{k,m}^{1},\bar{\delta}_{k,m}^{2}\right)$} by linear interpolation over $\mathcal{G}_k$ (given $\left(\delta_{k-1,m}^{1},\delta_{k-1,m}^{2}\right)$ if $\varepsilon>0$)
				\ENDFOR
				\STATE{compute the hedging error $-{\varphi\left( {S}^1_{\tau_{n,m},m} \right)}+(\bar{\delta}\cdot S_m)_{n}^{\tau_{n,m}/\Delta t}+C_{n}^{\tau_{n,m}/\Delta t}(\bar{\delta}) $} 
				\ENDFOR
				\STATE{estimate $\bar{p}_n=\rho\left(-\varphi\left(S^1_T \right)+\left(\bar{\delta}\cdot S\right)_{n}^{\tau_{n}/\Delta t} +C_{n}^{\tau_{n}/\Delta t}(\bar{\delta}) \right)e^{-r\left( \tau_n-t_n\right) }$ from simulations}
				\ENDWHILE
				\ENDFOR
				\IF{$n>0$}
				\STATE{compute and store coefficients for linear interpolation of $\bar{\delta}^1_n, \bar{\delta}^2_n$ and $\bar{p}_n$ over $\mathcal{G}_n$}
				\ENDIF
				\ENDFOR
				
				\STATE{Simulate $M_1$ random paths $\left(  {S}^1_{n,m}, {S}^2_{n,m}, {I}_{n,m},  {V}_{n,m}\right) $ for $n=1,\dots,N$} 
				\FOR{$m=1,\dots,M$ (for each simulated path)}
				\STATE{compute worst hedging stopping time $\tau_{n,m}$ by comparing payoff and hedging cost $\bar{p}_n$ (estimated by interpolation)}
				\FOR{$k=n+1,\dots,\tau_{n,m}/\Delta t$}
				\STATE{ compute the hedging strategy $\left(\bar{\delta}_{k,m}^{1},\bar{\delta}_{k,m}^{2}\right)$} by linear interpolation over $\mathcal{G}_k$ (given $\left(\bar{\delta}_{k-1,m}^{1},\bar{\delta}_{k-1,m}^{2}\right)$ if $\varepsilon>0$)
				\ENDFOR
				\STATE{compute the hedging error $-{\varphi\left( {S}^1_{\tau_{n,m},m} \right)}+(\bar{\delta}\cdot  {S}_m)_{n}^{\tau_{n,m}/\Delta t}+C_{n}^{\tau_{n,m}/\Delta t}(\bar{\delta}) $} 
				\ENDFOR
				\IF{$\varepsilon>0$}
				\RETURN{estimate $\bar{p}_0=\rho\left(-{\varphi\left( {S}^1_T \right)}+\left(\bar{\delta}\cdot  {S}\right)_{0}^{\tau_0/\Delta t}+C_{0}^{\tau_{0}/\Delta t}(\bar{\delta}) \right)e^{-r\tau_0}$}
				\ELSE
				\RETURN{estimate $\bar{p}_0=\rho\left(- {\varphi\left(S^1_{\tau_{0}} \right)}+\left(\bar{\delta}\cdot  {S}\right)_{0}^{\tau_{0}/\Delta t}\right)e^{-r\tau_0}$ and, for $n=1,\dots,N-1$, the quantiles of $S^1_{t_n},I_{t_n},\bar{\delta}^1_n$ and $\bar{\delta}^2_n$} based on the out of sample simulations
				\ENDIF
		\end{algorithmic}}
	\end{algorithm} 
	
\end{document}